\documentclass[conference]{IEEEtran}
\IEEEoverridecommandlockouts

\usepackage{bbding}
\usepackage{bbm}
\usepackage{graphicx}
\usepackage[cmex10]{amsmath}
\usepackage{cases}
\usepackage{amssymb}
\usepackage{amsfonts}
\usepackage{mathrsfs}
\usepackage{algorithmic}
\usepackage{array}
\usepackage{mdwmath}
\usepackage{eqparbox}
\usepackage[tight,footnotesize]{subfigure}
\usepackage[caption=true]{caption}
\usepackage[font=footnotesize]{subfig}

\usepackage{enumerate}
\ifx\pdfoutput\undefined
\usepackage[dvips]{color}
\else
\usepackage{color}
\fi
\usepackage{setspace}
\usepackage{cases}
\usepackage{algorithmic}
\usepackage{subfigure}
\usepackage{graphicx}
\usepackage{amsmath}
\usepackage{graphics}
\usepackage{epsfig}
\usepackage{fixmath}
\usepackage[ruled,vlined]{algorithm2e}

\ifCLASSINFOpdf
\else
\fi

\begin{document}

\title{Competitive Charging Station Pricing for Plug-in Electric Vehicles}

\author{Wei Yuan, \emph{Member, IEEE},
Jianwei Huang$^*$, \emph{Fellow, IEEE},\\
and Ying Jun (Angela) Zhang, \emph{Senior Member, IEEE}
\thanks{
The work is supported by a grant from the Research Grants Council of the Hong Kong Special Administrative Region, China, under Theme-based Research Scheme through Project No. T23-407/13-N, and the National Natural Science Foundation of China (Grant No. 61300223 and No. 61571205).

Part of the results in this paper were presented in IEEE SmartGridComm 2014 \cite{Conferencepaper}.

W. Yuan is with the School of Electronic Information and Communications, Huazhong University of Science and Technology, Wuhan, 430074, China. e-mail: yuanwei@mail.hust.edu.cn. J. Huang (corresponding author) and Y. J. Zhang are with the Department
of Information Engineering, The Chinese University of Hong Kong, Shatin, New Territories, Hong Kong. e-mail: \{jwhuang,yjzhang\}@ie.cuhk.edu.hk.

}
}

\maketitle

\begin{abstract}
This paper considers the problem of charging station pricing and plug-in electric vehicles (PEVs) station selection. When a PEV needs to be charged, it selects a charging station by considering the charging prices, waiting times, and travel distances. Each charging station optimizes its charging price based on the prediction of the PEVs' charging station selection decisions and the other station's pricing decision, in order to maximize its profit. To obtain insights of such a highly coupled system, we consider a one-dimensional system with two competing charging stations and Poisson arriving PEVs. We propose a multi-leader-multi-follower Stackelberg game model, in which the charging stations (leaders) announce their charging prices in Stage I, and the PEVs (followers) make their charging station selections in Stage II. We show that there always exists a unique charging station selection equilibrium in Stage II, and such equilibrium depends on the charging stations' service capacities and the price difference between them. We then characterize the sufficient conditions for the existence and uniqueness of the pricing equilibrium in Stage I. We also develop a low complexity algorithm that efficiently computes the pricing equilibrium and the subgame perfect equilibrium of the two-stage Stackelberg game.
\end{abstract}

\IEEEpeerreviewmaketitle

\section{Introduction}

Since the sales of the first highway-capable all electrical vehicle Tesla Roadster in 2008, there have been more than 290,000 similar plug-in electrical vehicles (PEVs) sold in the US as of December 2014. As PEVs help reduce the emission of greenhouse gas, there has been a growing interest from both industry and academia in terms of the technology and economics aspects of PEV deployment ~\cite{Conferencepaper}-\cite{PHEV4}.

One important limitation for the PEV is its limited battery capacity. A wide deployment of PEVs requires an extensive charging station network, which fortunately is being deployed in many countries. For example, today there are more EV charging points than gas stations in Japan~\cite{Japan}. In the US, operators such as CarCharging provide national-wide public PEV charging services. As the number of charging operators in the electrical vehicle market increases, the issue of competitive pricing among charging stations is getting increasingly important and practical. For example, in some Chinese cities including Beijing and Qingdao, PEV owners have to pay service fees in addition to electricity bills to charge their cars. The government declares the maximally allowed service fee, and the operators such as Tellus Power and Potevio set their own charging prices (i.e., the sum of electricity price and service price) to maximize their own revenue subject to the government rules.

The owners of the PEVs, on the other hand, are able to identify multiple close-by charging stations using mobile applications such as PlugShare~\cite{plugshare}, and choose those that offer the best cost and distance tradeoff. The interactions between the charging stations and (owners of) PEVs will determine the dynamics and equilibrium of such a charging market.

In this work, we aim to answer the following key research questions:

$\bullet$ How should a PEV select a charging station based on the charging prices, travel distances, and the expected waiting times of multiple close-by stations?

$\bullet$ How should a charging station optimize its charging price to maximize its profit, considering the decisions of the competing charging stations and the responses of the PEVs?

Addressing these questions are challenging due to the tight coupling, among different PEVs, among multiple charging stations, and between PEVs and charging stations.

To shed some insights on this complicated problem, we consider a stylized one-dimensional system model, with two competing charging stations and dynamically arriving PEVs. The charging stations announce their charging prices simultaneously at the beginning of a day, and the PEVs make their selections asynchronously during the day as their charging needs arise. We formulate the problem as a \emph{multi-leader-multi-follower} Stackelberg game~\cite{stackelberg game}, in which the charging stations are the leaders making decisions in Stage I, and the PEVs are the followers making decisions in Stage II. We then characterize the charging stations' pricing and the PEVs' selection behaviors at the equilibria of this game.

Our main contributions are summarized as follows.

$\bullet$ \emph{Novel and Practical Model:} To the best of our knowledge, this is the first work that jointly studies competitive charging station pricing and PEV station selection. Our model considers the heterogenous service capabilities and asymmetric locations of charging stations. It also captures a PEV's waiting time before it receives the charging service, hence is different from existing PEV station selection models in the literature.

$\bullet$ \emph{Analysis and Insights:} Under any fixed charging station prices, we show that there always exists a unique PEV station selection equilibrium. We further show that such an equilibrium can be one of five types (three \emph{pure-strategy} ones and two \emph{mixed-strategy} ones), depending on the prices and the service capacities. Then we characterize the sufficient conditions for the existence and uniqueness of charging station pricing equilibrium, considering the PEVs' station selection behaviors. Our analysis suggests that a charging station with a position advantage usually declares a higher equilibrium price than its competitor.

$\bullet$ \emph{Efficient Algorithm:} We propose a low complexity algorithm to compute the pricing equilibrium, which may not require explicit information exchange between the charging stations and is provably convergent.

The remainder of this paper is organized as follows. Section II discusses the related work. Section III presents the system model. We analyze the equilibria for our proposed games in Sections IV and V, respectively. In Section VI, we propose the algorithm to compute the pricing equilibrium. Numerical results are provided in Section VII, followed by the concluding remarks in Section VIII.

\section{Related Work}
\subsection{ Pricing behaviour of charging stations}
Existing literature on multi-station pricing models can be divided into two categories depending on the relationship among charging stations: non-competitive and competitive.

\subsubsection{Non-competitive pricing}

Bayram \emph{et al.} in  \cite{multi seller 2} assumed that all charging stations have the same charging prices, and proposed a method to optimize the prices to maximize the sum of the utilities of all charging stations. In \cite{revieweradd 1}, Bayram \emph{et al.} proposed a single-leader-multi-follower Stackelberg game, in which the charging network operator acts as the leader and the EV customers are the followers. The leader optimizes the prices of charging stations, and each follower makes a selection between the nearest station and a less busy one. Ban \emph{et al.} in \cite{PHEV6} investigated the PEV allocation and dynamic price control for multiple charging stations. Different from \cite{multi seller 2}, the authors in \cite{PHEV6} considered heterogeneous prices. Assuming that each PEV prefers the charging station with the lowest price, the authors of \cite{PHEV6} proposed a price control scheme to implement optimal PEV allocation.

\subsubsection{Competitive pricing}
Escudero-Gar$\acute{z}$as and Seco-Granados in \cite{PHEV5} studied the competitive pricing of multiple charging stations using the Bertrand's oligopoly model. The demand for each charging station was assumed to be a linearly decreasing function of its price. Lee \emph{et al.} in  \cite{competitive pricing} assumed that the EVs are uniformly distributed and select their charging station based on the prices, the distances, and the preferences. Furthermore, every charging station is able to generate electricity and can sell its electricity to the power grid and the EVs. The authors of \cite{competitive pricing} formulated the price competition among multiple charging stations as a supermodular game model. A key assumption of \cite{competitive pricing} is that the choices of PEVs are independent of each other, which is not the case in our model.

\subsection{Hotelling game model}
Our model is related to the hotelling game \cite{hotelling 1}, where two geographically separated sellers compete to serve customers at different locations. However, most hotelling game models cannot be directly used to understand the interplay between charging stations and PEVs, since they do not consider the dependence among the customers' decisions \cite{hotelling 1}-\cite{hotelling 3}. Gallay and Hongler in \cite{hotelling 4} extended the hotelling game by introducing the waiting time cost, hence the customers' decisions are coupled due to the consideration for waiting time. However, the model in \cite{hotelling 4} did not consider the competition among stations and assumed that all stations announce the same price.

\subsection{Summary}

This work differs from the existing results in two respects.
\begin{enumerate}
\item In our system, the PEVs take into account their expected waiting times when selecting charging stations. The waiting time is non-negligible for the PEVs due to their relatively long charging times \cite{revieweradd 2}\cite{revieweradd 3}. This is a key difference between our work and the existing models developed for gasoline stations \cite{gasoline 1} \cite{gasoline 2}, where the queueing delay at the station is considerably shorter.
\item Our Stackelberg game (i.e., leader-follower game) jointly studies the PEVs' station selecting and the charging stations' pricing decisions, instead of treating only one of them in the literature. More specifically, we consider two leaders (i.e., charging stations) who choose prices to optimize their own revenues. This is different from the single-leader model studied in \cite{revieweradd 1}\cite{revieweradd 2}\cite{revieweradd 3}.
\end{enumerate}

Due to the high complexity of solving the general form of the considered problem, here we focus on a simple one-dimensional system with two stations. The insights from the work can help us understand more general systems involving more than two stations with more general locations.

\section{Problem Formulation and Game Model}

\begin{figure}[h]
\centering
        \scalebox{0.5}[0.5]{\includegraphics {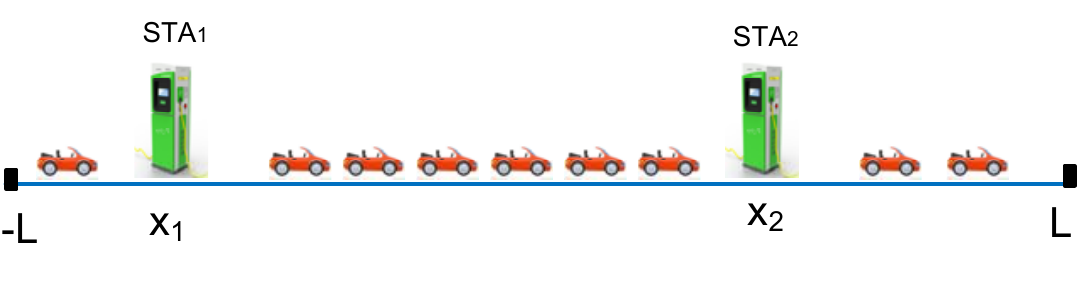}}
        \caption{The considered system.}
\end{figure}

As shown in Fig. 1, our system is represented by a line segment, denoted by $[-L,L]$. The line
model also applies to the scenario where both stations are along a zigzag or curving road without major branches
within a significant distance. The charging stations, which belong to different operators, are located at $x_1$ and $x_2$ ($-L<x_1<x_2<L$), respectively\footnote{In practice, the charging stations in the same small region may be controlled by different operators. In Frisco, Texas, USA, for example, the distance between two charging stations owned by two different operators (Voltaic Electrical and Blink Network) can be shorter than 10 Kilometers.  }. We allow the two stations to be asymmetrically located on the line, \emph{i.e.}, $x_1 \neq -x_2$.

Each charging station serves its customers (\emph{i.e.}, the PEVs waiting in its queue) on a first come first served basis.\footnote{In this work, we do not consider the stations with a battery replacement strategy \cite{revieweradd new}. } Suppose that charging station $i\in \{1,2\}$ has $k_i\geq 1$ identical charging ports, and the service (charging) time for a PEV follows a general distribution with a mean $\frac{1}{\mu_i}$ ($\mu_i>0$ is the average charging rate) and a variance $\sigma_i^2$. Accordingly, we can use $k_i\mu_i$ to measure the service capacity of charging station $i$. Without losing of generality, we assume that $k_1\mu_1\geq k_2\mu_2$. In practice, it is possible to have $\mu_1 \neq \mu_2$ if the charging stations use different levels of charge specifications~\cite{PHEV5} or different chargers (such as AC Level 2 charger and DC Fast charger).

The system works in two stages:

$\bullet$ In Stage I, two charging stations simultaneously determine their charging prices at the beginning of a time period\footnote{Different time periods can have different PEV arrival rates and electricity prices.}, and periodically broadcast the (fixed) prices together with the stations' locations to the potential customers (PEVs) throughout the day (\emph{e.g.}, through a mobile app such as PlugShare).

$\bullet$ In Stage II, given the charging prices and the locations of both charging stations, every PEV independently selects a charging station to recharge its battery. We assume that the PEV driver does not have any prior preference regarding which direction he would drive on the road.

Note that the decisions among the PEVs, among the charging stations, and between PEVs and charging stations are actually interdependent. To characterize the interplay of PEVs and charging stations, we propose a two-stage Stackelberg model, which consists of two games at two different levels:

$\bullet$ Charging Station Pricing Game (CSPG): In Stage I, the charging stations engage in a \emph{CSPG} game, in which they determine their charging prices by considering each other's pricing choices and the PEVs' selection choices in Stage II.

$\bullet$ PEV Station Selection Game (PSSG): In Stage II, the PEVs engage in a \emph{PSSG} game, in which each PEV selects its charging station under fixed prices from the charging stations, by considering the station choices of other PEVs.

With this hierarchical game, we aim to derive a stable decision outcome for charging stations and PEVs.

\subsection{PEV Station Selection Game in Stage II}
The \emph{PSSG} game is defined as follows:
\begin{itemize}
\item Players: The PEVs.
\item Strategies: The strategy of a PEV $n$ corresponds to its station choice $s_n(x)\in\{1,2\}$, where $x$ is the location of PEV $n$. 
\item Payoffs: The payoff of a PEV is the negative value of its cost, which includes three parts: traveling cost, waiting time, and charging cost.
\end{itemize}

\subsubsection{Traveling Cost}
We assume that the there is no traffic jam and the traveling delay is determined by the traveling distance only. Let $l_{n,s_n}$ denote the distance from the current location of PEV $n$ (\emph{i.e}., $x$) to the selected charging station $s_n$. Then we have $l_{n,s_n}=|x-x_{s_n}|$. The traveling cost of PEV $n$ is $k_l l_{n,s_n}$, where $k_l>0$.
\subsubsection{Charging Cost}
For simplicity, we assume all PEVs have the same battery capacity, start to charge when their battery is close to empty (or to the same level of charging level suggested by the manufacturer), and want to get fully charged before leaving the station. Hence all PEVs have the same demand\footnote{Another way to understand the homogeneous demand assumption is to consider the charging station's perspective. Since the charging station does not have complete information of the PEVs, then it will be difficult for the station to predict the precise demand of each PEV. As considering the M/G/k queueing model used in this paper is already complicated enough, we simply use the average demand of the PEVs to represent the demand of each PEV, so that the charging station can compute the competitive charging price.}.  We use $d$ to denote charging demand of an PEV, and $p_{s_n}$ to denote the price of the selected charging station. The charging cost is $k_p d p_{s_n}$, where $k_p>0$.
\subsubsection{Waiting time}
Now we estimate the waiting time using queueing theory. The inter-arrival time between two consecutive PEVs arriving at the same charging station depends on two factors: 1) the time interval between the time instances at which these two PEVs decide to seek charging service,  and 2) the difference between the travel times to the station. Since a PEV will only consider charging stations close-by, we assume that the difference between the travel times to these two stations are negligible\footnote{Take charging station 1 as an example. The difference between the travel times of two PEVs is no more than $\frac{L-x_1}{v}$, where $v$ is the speed of PEVs. If $v$ equals to 60 kph and $|L-x_1|$ is 5 km, it will be no more than 5 mins. On the other hand, if a PEV is charged on a standard 120-volt outlet, it usually needs 8 hours to be fully charged. If the PEV uses a dedicated 240 circuit, it may need 3 hours. If the PEV uses a 480V circuit, it needs 20 to 30 mins~\cite{charging time}.}. We further assume that the time interval between the charging decisions of two consecutive PEVs at the same location (the time instances when they decide seeking charging services) is exponentially distributed. If all the PEVs generated from this location select the same charging station, then the time interval between the arrivals of two consecutive PEVs at the station is exponentially distributed. Accordingly, we can consider the PEV stream generated from this location as Poisson arrivals, or a Poisson stream~\cite{revieweradd new2013}\cite{multi seller 2}\cite{single seller 3}-\cite{Poisson 2}. We also assume that the PEV streams are uniformly distributed in the entire one-dimensional system. Then, we consider the stream of PEVs from any line segment as an \emph{aggregation} of PEV streams, which is also Poisson \cite{Poisson aggregation}.

Suppose that the arrival rate of the PEVs coming from a unit line segment is $\lambda>0$. Let $A_{i}\subseteq [-L,L]$ be the set of locations of the PEVs who select charging station $i$. In general, $A_i$ might include multiple disjoint segments. We use $|A_{i}|$ to denote the total length of $A_{i}$. Then, the arrival rate of the PEVs selecting charging station $i$ is $|A_i|\lambda$.

Due to the small penetration of PEVs in today's market, $\lambda$ is usually small and the probability that a station having no waiting room for a PEV will be very small. For simplicity, here we employ the theory of $M/G/k$ queue~\cite{queueing theory} to estimate the waiting time of a PEV at a station.\footnote{It should be pointed out that the M/G/K model is a \emph{statistic} model, and it does not reflect the actual waiting time realization experienced by a certain PEV.} Then the mean waiting time of a PEV at charging station $i$ is given by
\begin{equation}\label{waiting time}
q_i(|A_i|) \approx\frac{|A_{i}|\lambda (\sigma^2_{i}+\frac{1}{\mu_{i}^2})\rho_{i}^{k_{i}-1}}{2(k_{i}-1)!(k_{i}-\rho_{i})^2\large[\sum_{m=0}^{k_{i}-1}\frac{\rho_{i}^m}{m!} +\frac{\rho_{i}^{k_{i}}}{(k_{i}-1)!(k_{i}-\rho_{i})}\large]}
\end{equation}
where $\rho_{i}=\frac{|A_{i}|\lambda}{\mu_{i}}$. In \eqref{waiting time}, $k_i$, $\mu_i$ and $\sigma_i$ are the parameters of charging stations, and $\lambda$ is the parameter of PEVs.

 Using the above notations, the payoff of PEV $n$ is defined as
\begin{equation}\label{utility_pev}
U_n(\boldsymbol{s})=-k_l l_{n,s_n}-k_q q_{s_n}(\boldsymbol{s})-k_p d p_{s_n}
\end{equation}
where $\boldsymbol{s} = \{s_n,n\in\mathcal{N}\}$ is the strategy profile of all PEVs (which can also be captured by $A_1$ and $A_2$), $\mathcal{{N}}$ is the set of all PEVs, and $k_q>0$.

Since there are no direct communications between PEVs, we assume that a PEV does not know other PEVs' decisions when making its decision. Hence, although the PEVs make charging station selection asynchronously during the day, we can model PSSG as a \emph{simultaneous move} (or static) game\footnote{Even though the decisions may be made at different time instances, the game is ``simultaneous" because each player has no information about the decisions made before or after his.}.

\subsection{Charging Station Pricing Game in Stage I}
The \emph{PSSG} game is defined as follows:
\begin{itemize}
\item Players: The charging stations.
\item Strategies: The strategy of a charging station $i$ is its price $p_i$.
\item Payoffs: The payoff of a charging station includes two parts: 1) the revenue of providing charging service to the PEVs, and 2) the fixed and operational costs.
\end{itemize}

In \emph{CSPG}, the charging stations simultaneously determine their charging prices $p_1$ and $p_2$. We assume that $p_i \in [p_{\min},p_{\max}]$, and the unit electricity cost paid by a charging station $i$ to the utility company is $c_i\leq p_{\min}$. Charging station $i$ has a fixed cost $\check{c}_i$ for providing the service (such as labour cost), which is assumed to be independent of the number of PEVs requesting the service.
The payoff of charging station $i$ can be written as
\begin{equation}
Q_i(p_i,p_{j})=(p_i - c_i)D_i(p_i,p_{j})-\check{c}_i.
\end{equation}
Here $D_i(p_i,p_{j})=|A_i|d$ is the total demand of the PEVs selecting charging station $i$, which depends on the prices of both stations.

Next we will derive the subgame perfect equilibrium (SPE) of the Stackelberg game, which represents a Nash Equilibrium (NE) of every subgame of the game. Under an SPE, neither PEVs nor charging stations have incentives to change their strategies. Hence SPE corresponds to the stable pricing and station selection outcome. We use \emph{backward induction} to derive the SPE. More specifically, we start with Stage II (\emph{PSSG}) and analyze the PEVs' selection given prices $p_1$ and $p_2$. Then, we look at Stage I (\emph{CSPG}) and analyze how charging stations make the pricing decisions, taking the PEVs' responses in Stage II into consideration.

\section{PEV Station Selection Game in Stage II}

Under a given price pair $(p_1,p_2)$, every PEV selects a charging station to maximize its payoff in \eqref{utility_pev}. If too many PEVs select the charging station $i$ such that $\frac{|A_i|\lambda}{k_i\mu_i}\geq 1$, the queue at charging station $i$ grows to infinity. Clearly, the PEVs will try to avoid overloading a charging station due to the concern about the waiting time. However, when $k_1\mu_1+k_2\mu_2\leq 2L\lambda$ holds, overload is inevitable since the total serving capacity of two stations is not enough to serve all the PEVs. In this situation, one should increase the number of ports (\emph{i.e.}, $k_i$) or introduce better (faster) charging technologies (e.g., DC Fast charging). In this paper, we will consider the more meaningful case of $k_1\mu_1+k_2\mu_2>2L\lambda$, which means that by appropriately selecting their stations, the PEVs can avoid overloading any charging station.

The above discussions reveal that the service capacity of a charging station, \emph{i.e.}, $k_i\mu_i$, has a significant impact on the PEVs' decisions. For example, if $\frac{2L\lambda}{k_i\mu_i}$ approaches $1$, some PEVs must select charging station $j \neq i$ to avoid overloading charging station $i$. For the convenience of later discussions, we can classify different levels of a charging station's service capacity in terms of FULL, HIGH, MIDDLE and LOW.

\begin{figure}[h!]
\centering
        \scalebox{0.55}[0.55]{\includegraphics {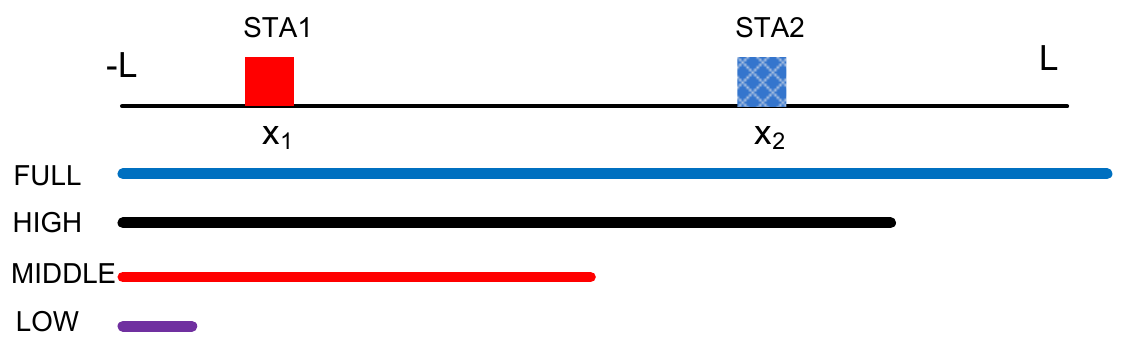}}
        \caption{ Levels of service capacity of charging station 1. The length of the line in each case illustrates the maximum number of PEVs that station 1 can serve.}
\end{figure}

Take charging station 1 for example. As shown in Fig. 2, FULL means that station 1 alone can serve all the PEVs in the system (along the entire segment of length $2L$) without being overloaded. HIGH, MIDDLE, and LOW indicate that the station can serve the PEVs in a segment with a length $x+L$, where $x_2<x<L$ for HIGH,  $x_1<x<x_2$ for MIDDLE and $-L<x<x_1$ for LOW. Such a terminology allows us to classify the system model into nine scenarios\footnote{It should be pointed out that the scenarios LOW-LOW and MIDDLE-LOW do not exist in a stable system, as the charging stations cannot serve all the PEVs in these scenarios.}:
\begin{enumerate}
\item FULL-FULL: $k_1\mu_1>2L\lambda$ and $k_2\mu_2>2L\lambda$.
\item  FULL-HIGH: $k_1\mu_1>2L\lambda$ and $(L-x_1)\lambda<k_2\mu_2\leq 2L\lambda$.
\item FULL-MIDDLE: $k_1\mu_1>2L\lambda$ and $(L-x_2)\lambda<k_2\mu_2\leq (L-x_1)\lambda$.
\item FULL-LOW: $k_1\mu_1>2L\lambda$ and $k_2\mu_2\leq (L-x_2)\lambda$.
\item HIGH-HIGH: $(L+x_2)\lambda<k_1\mu_1\leq 2L\lambda$ and $(L-x_1)\lambda<k_2\mu_2\leq 2L\lambda$.
 \item  HIGH-MIDDLE: $(L+x_2)\lambda<k_1\mu_1\leq 2L\lambda$ and $(L-x_2)\lambda<k_2\mu_2\leq (L-x_1)\lambda$.
 \item HIGH-LOW: $(L+x_2)\lambda<k_1\mu_1\leq 2L\lambda$ and $k_2\mu_2\leq (L-x_2)\lambda$.
 \item  MIDDLE-HIGH: $(L+x_1)\lambda<k_1\mu_1\leq (x_2+L)\lambda$ and $(L-x_1)\lambda<k_2\mu_2\leq 2L\lambda$.
\item  MIDDLE-MIDDLE: $(L+x_1)\lambda<k_1\mu_1\leq (x_2+L)\lambda$ and $(L-x_2)\lambda<k_2\mu_2\leq (L-x_1)\lambda$.
\end{enumerate}

As an example, let us look at the FULL-HIGH scenario. In this scenario, charging station 1 has the full service capacity and can serve all the PEVs. Charging station 2 has a limited service capacity, specified by $(L-x_1)\lambda<k_2\mu_2\leq 2L\lambda$. Therefore, it can serve only some of PEVs in the system due to $|A_2|<\frac{k_2\mu_2}{\lambda}<2L$. For example, it is able to serve the PEVs in the segment $(L-\frac{k_2\mu_2}{\lambda},L]$ alone.

Since there are many PEVs in this model, we can view the Stage II game as a population game \cite{population}, where a single PEV's selection change will not affect the lengths of sets $A_1$ and $A_2$ and the corresponding waiting times at the two stations. Let us use $U_n(s_n;A_1,A_2)$ to denote PEV $n$'s payoff, where other PEVs' station choices are represented by sets $A_1$ and $A_2$. To maximize its payoff, a PEV compares the payoffs of selecting two charging stations. If $U_n(i;A_1,A_2)\geq U_n(j;A_1,A_2)$, PEV $n$ will select charging station $i$ instead of $j$. When no PEV has an incentive to change its selection unilaterally, a stable decision outcome emerges. Such outcome corresponds to the NE of $PSSG$, which is define as follows.

\newtheorem{definition}{\textbf{Definition}}
\begin{definition}
A strategy profile $\boldsymbol{s}^\ast=\{s^*_n, \forall n \in\mathcal{N}\}$ is an NE of the \emph{PSSG} if $U_n(s_n^*;A_1,A_2)\geq U_n(s^\prime_n;A_1,A_2)$ for every $n\in\mathcal{N}$, where $s'_n$ is the different charging station selection than $s_n^\ast$.
\end{definition}

 We will show that there exist at most five types of NEs in \emph{PSSG}, \emph{i.e.}, three pure-strategy NEs and two mixed-strategy NEs, as shown in Fig. 3. Which type of NE will emerge depends on the price difference between the charging stations and the type of system model scenario.

\begin{figure}[h!]
\centering
        \scalebox{0.30}[0.3]{\includegraphics {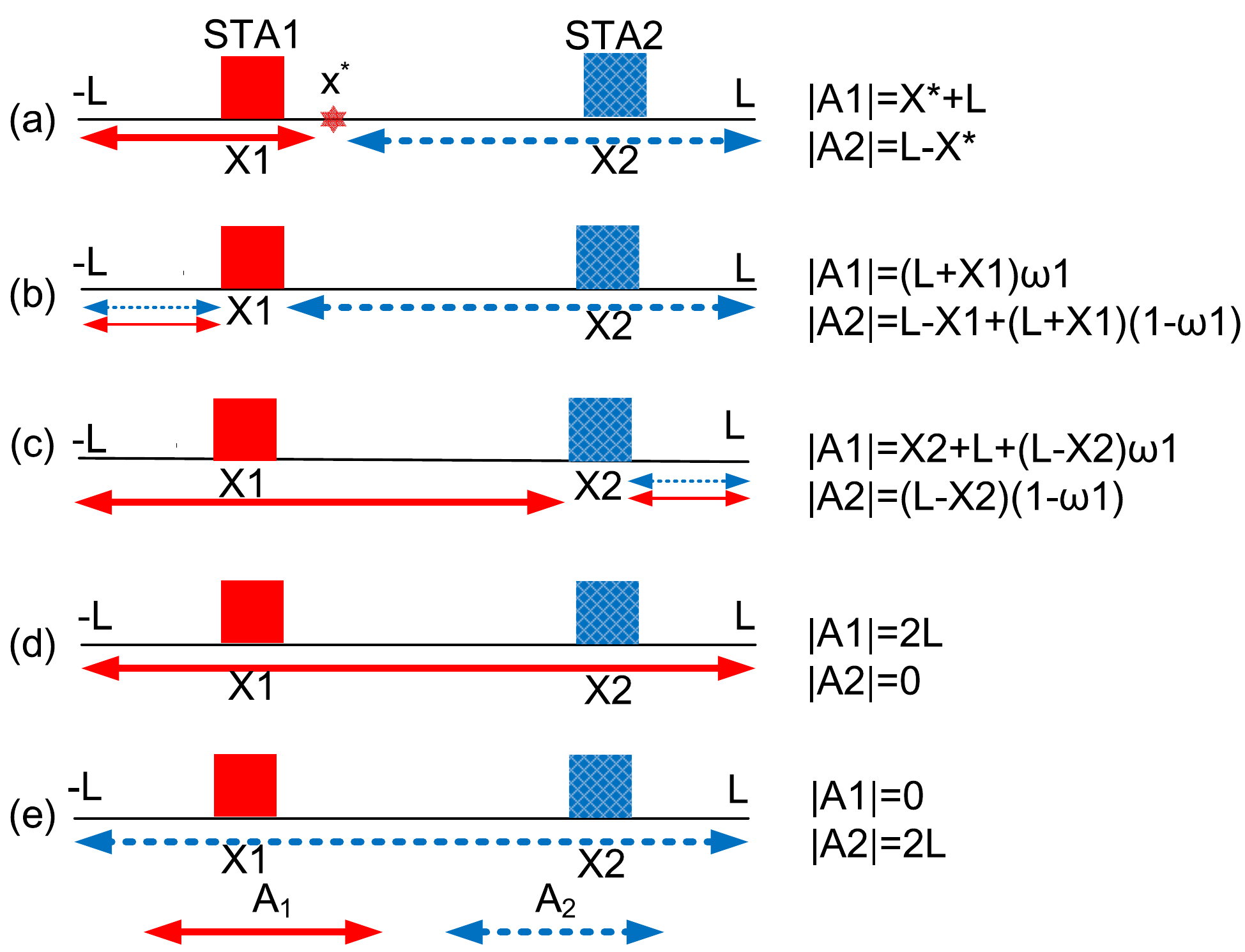}}
        \caption{Five types of NEs in \emph{PSSG}. Here red segments correspond to $A_1$ and blue segments correspond to $A_2$. When red and blue segments overlap, we have a mixed strategy equilibrium. }
\end{figure}

\subsection{The FULL-FULL Scenario}
In the FULL-FULL scenario, all five types of NEs can emerge under different prices.

\subsubsection{Pure-strategy NE with an indifference point}
 In most cases, every PEV has one preferred charging station. According to \eqref{utility_pev}, under given $A_1$ and $A_2$, a PEV determines its preferred charging station based on its location and the price difference $p_1-p_2$. We may find a particular location, where a PEV at this location is indifferent of selecting between the two charging stations, \emph{i.e.}, $U_n(1;A_1,A_2)=U_n(2;A_1,A_2)$. We call such location the \emph{indifference point}, denoted by $x^*$.

\newtheorem{theorem}{\textbf{Theorem}}
 \newtheorem{lemma}{\textbf{Lemma}}

The existence of an indifference point actually depends on the price difference $p_1-p_2$. Intuitively, the indifference point may not exist if $|p_1-p_2|$ is too high (in which case all PEVs prefer the same charging station). When an indifference point exists, we have a pure-strategy charging station selection NE.

To characterize the conditions for such a pure-strategy NE, we define two thresholds as follows,
\begin{equation*}
\theta^L_1=-\frac{k_q(q_1(L+x_2)-q_2(L-x_2))+k_l(x_2-x_1)}{k_p\cdot d},
\end{equation*}
and
\begin{equation*}
\theta^R_1=\frac{k_q(q_2(L-x_1)-q_1(x_1+L))+k_l(x_2-x_1)}{k_p\cdot d}.
\end{equation*}

Then we have the following theorem.
\begin{theorem}
In the FULL-FULL scenario, if $p_1-p_2 \in (\theta^L_1,\theta^R_1)$, then the NE strategy of PEV $n$ at a location $x \in [x_1,x_2]$ is
\begin{equation}
 s_n^\ast(x) =
  \begin{cases}
   1, &\mbox{if $x\in[-L, x^\ast)$},\\
   2, &\mbox{if $x \in [x^\ast, L]$},
   \end{cases}
\end{equation}
where $x^*$ is the unique root of
\begin{multline}\label{calculate_indifference}
k_pd(p_1-p_2)+k_l(2x-x_1-x_2)\\
+k_q (q_1(x+L)-q_2(L-x))=0.
\end{multline}
\end{theorem}

The proof of Theorem 1 is given in Appendix A. At the NE described in Theorem 1, all the PEVs on the left side of the indifference point select charging station 1, and the rest PEVs select charging station 2, as shown in Fig. 2 (a). In other words, the one-dimensional system can be divided into two continuous segments, each of which is served by one charging station.

\subsubsection{Mixed-strategy NE}
By analyzing \eqref{calculate_indifference}, we can show that $x^\ast$ decreases with $p_1-p_2$. As $p_1-p_2$ increases, $x^*$ will move closer to $x_1$, and charging station 2 will attract more PEVs. Once $p_1-p_2=\theta^R_1$, the indifference point $x^{\ast}=x_1$.
In this case, charging station 2 attracts all the PEVs on the right side of station 1 at the NE.

Once the price difference $p_1-p_2>\theta^R_1$, a new type of equilibrium emerges and it is no longer a pure-strategy NE. At this equilibrium, some PEVs in the range of $[-L, x_1]$ select charging station 1, while other PEVs in the same range $[-L,x_1]$ select charging station 2. In fact, for a PEV in the range of $[-L,x_1]$, whether selecting station 1 or station 2 no longer depends on its location; it only depends on the values of $|A_1|$ and $|A_2|$. We illustrate such an equilibrium in Fig. 3 (b), where the fraction of PEVs in the range of $[-L,x_1]$ selecting charging station 1 (denoted by the red segment) needs to be properly chosen, such that $U_n(1;A_1,A_2)=U_n(2;A_1,A_2)$ for all PEVs in this range, which corresponds to a mixed-strategy equilibrium. Similarly, when $p_1-p_2<\theta^L_1$, we have another mixed strategy equilibrium, as shown in Fig. 3 (c).

More formally, under a mixed-strategy equilibrium, some PEV will randomly select two strategies with some positive probabilities for both. Let $\omega_1(x)$ and $1-\omega_1(x)$ denote the probability of the PEV at a location $x$ selecting charging stations 1 and 2, respectively. Then we use $(\omega_1(x),1-\omega_1(x))$ to represent the mixed-strategy of such a PEV.

 Before characterizing the mixed-strategy NE, we further define two more thresholds,
\begin{equation*}
\theta^L_2=-\frac{k_qq_1(2L)+k_l(x_2-x_1)}{k_p\cdot d},
\end{equation*}
and
\begin{equation*}
\theta^R_2=\frac{k_qq_2(2L)+k_l(x_2-x_1)}{k_p\cdot d}.
\end{equation*}

\begin{theorem}
(1) In the FULL-FULL scenario, if $p_1-p_2 \in [\theta^R_1,\theta^R_2)$, then the NE strategy of PEV $n$ at a location $x$ is
 \begin{equation}
 s_n^\ast(x) =
  \begin{cases}
   {\color{black}{(0, 1)}}, &\mbox{if $x\in[x_1, L]$},\\
   (\omega_1,1-\omega_1), &\mbox{if $x \in [-L, x_1)$},
   \end{cases}
\end{equation}
 where $\omega_1$ is the unique root of the following equation in the range of $[0,1]$,
    \begin{multline}\label{probability_indifference}
        k_q(q_1((x_1+L)\omega_1)-q_2(L-x_1+(x_1+L)(1-\omega_1)))\\
        +k_pd(p_1-p_2)+k_l(x_1-x_2)=0.
    \end{multline}

(2) In the FULL-FULL scenario, if $p_1-p_2 \in (\theta^L_2,\theta^L_1]$, then the NE strategy of PEV $n$ at a location $x$ is
 \begin{equation}
 s_n^\ast(x) =
  \begin{cases}
    {\color{black}{(1, 0)}}, &\mbox{if $x\in[-L,x_2]$},\\
   (\omega_1,1-\omega_1), &\mbox{if $x \in (x_2, L]$},
   \end{cases}
\end{equation}
where $\omega_1$ is the unique root of the following equation in the range of $[0,1]$,
    \begin{multline}\label{probability_indifference 2}
        k_q(q_2((L-x_2)(1-\omega_1))-q_1(x_2+L+(L-x_2)\omega_1))\\
        +k_pd(p_2-p_1)+k_l(x_1-x_2)=0.
    \end{multline}
\end{theorem}
The proof of Theorem 2 is given in Appendix B.

\subsubsection{Pure-strategy NE with a dominant station}
From \eqref{probability_indifference} and \eqref{probability_indifference 2}, we can show that $\omega_1$ decreases with $p_1-p_2$ in both cases. When $p_1-p_2$ reaches the critical point $\theta^L_2$ (or $\theta^R_2$), all the PEVs select charging station 1 (or 2), as shown in Fig. 2 (d) (or Fig. 2(e)). This leads to a new type of pure-strategy NE, where all PEVs adopt the same strategy and choose the same ``dominant" station. Furthermore, if charging station 1 (or 2) keeps decreasing  below $\theta_2^L$ (or increasing above $\theta_2^R$), the selection outcome will remain unchanged, as shown by the following theorem.

\begin{theorem}
Consider the FULL-FULL scenario. If $p_1-p_2 \in [\theta^R_2,p_{\max}-p_{\min}]$, then the NE strategy for any PEV is $s_n^\ast(x)=2$. If $p_1-p_2 \in [p_{\min}-p_{\max},\theta^L_2]$, then the NE strategy of any PEV is $s_n^\ast(x)=1$.
\end{theorem}

The proof of Theorem 3 is given in Appendix C. According to Theorems 1 to 3, we can conclude that {\emph{PSSG}} always has a unique NE.

\subsection{The Other Eight Scenarios}

In the other eight scenarios, however, not all types of NEs can emerge due to the limited service capacity of at least one charging station. For illustration, here we consider two scenarios: HIGH-HIGH and MIDDLE-MIDDLE.

\subsubsection{HIGH-HIGH}
In this scenario, $k_i\mu_i\leq 2\lambda L$ for any $i\in \{1,2\}$, and neither station can serve all the PEVs in the system. Therefore, the NEs shown in Fig. (d) and Fig. (e) cannot be achieved, and there exist three types of NE in this scenario, as shown in Fig. 3(a), Fig. 3(b) and Fig. 3(c).
\begin{theorem}
(1) In the HIGH-HIGH scenario, if $p_1-p_2 \in (\theta^L_1,\theta^R_1)$, then the NE strategy of PEV $n$ at a location $x \in [x_1,x_2]$ is
\begin{equation}
 s_n^\ast(x) =
  \begin{cases}
   1, &\mbox{if $x\in[-L, x^\ast)$},\\
   2, &\mbox{if $x \in [x^\ast, L]$},
   \end{cases}
\end{equation}
where $x^*$ is the unique root of \eqref{calculate_indifference} and $x^*\in [x_1,x_2]$.\\
(2) In the HIGH-HIGH scenario, if $p_1-p_2\geq \theta^R_1$, then the NE strategy of PEV $n$ at a location $x$ is
 \begin{equation}
 s_n^\ast(x) =
  \begin{cases}
   {\color{black}{(0, 1)}}, &\mbox{if $x\in[x_1, L]$},\\
   (\omega_1,1-\omega_1), &\mbox{if $x \in [-L, x_1)$},
   \end{cases}
\end{equation}
 where $\omega_1$ is the unique root of \eqref{probability_indifference} and $\frac{2L\lambda-k_2\mu_2}{(L+x_1)\lambda}<\omega_1<1$.\\
 (3) In the HIGH-HIGH scenario, if $p_1-p_2 \leq \theta^L_1$, then the NE strategy of PEV $n$ at a location $x$ is
 \begin{equation}
 s_n^\ast(x) =
  \begin{cases}
    {\color{black}{(1, 0)}}, &\mbox{if $x\in[-L,x_2]$},\\
   (\omega_1,1-\omega_1), &\mbox{if $x \in (x_2, L]$},
   \end{cases}
\end{equation}
where $\omega_1$ is the unique root of \eqref{probability_indifference 2} and $0<\omega_1<\frac{k_1\mu_1-(L+x_2)\lambda}{(L-x_2)\lambda}$.
\end{theorem}
The proof of Theorem 4 is similar to the proofs of Theorems 1 and 2, and hence is omitted here.

\subsubsection{MIDDLE-MIDDLE}
In this scenario, the NEs shown in Fig. 3 (b), Fig. 3 (c), Fig. 3 (d), and Fig. 3 (e) cannot be achieved, as neither station has enough capacity (to serve the PEVs neighbouring its competitor). Therefore, there exists only one type of NE, as shown in Theorem 5.
\begin{theorem}
In the MIDDLE-MIDDLE scenario, the indifference point is \emph{unique} and the NE strategy of PEV $n$ at location $x$ is
\begin{equation}
 s_n^\ast(x) =
  \begin{cases}
   1, &\mbox{if $x \in [-L, x^\ast]$},\\
   2, &\mbox{if $x\in (x^\ast, L]$},
   \end{cases}
\end{equation}
where $x^\ast$ is the unique root of \eqref{calculate_indifference} and $L-\frac{k_2\mu_2}{\lambda}<x^{\ast}< \frac{k_1\mu_1}{\lambda}-L$.
\end{theorem}
The proof of Theorem 5 is similar to that of Theorem 1 and hence is omitted here.

\subsubsection{Summary}
The above equilibrium analysis method is also applicable to the other scenarios. Here we summarize the equilibrium analysis for them in Table I. It should be pointed out that our analysis is applicable to a more general system. In Appendices D and E, we show how to extend our low-level game to some more general scenarios.

\begin{table} \caption {NEs under various scenarios
} \centering

\begin{tabular}{|c|c|c|}
  \hline
  \textbf{Scenario} & \textbf{Possible NEs} & \textbf{Illustration}\\
   \hline\hline
  FULL-HIGH & 2 pure, 2 mixed & Fig. 3 (a),(b),(c),(d) \\
   \hline
  FULL-MIDDLE &  2 pure, 1 mixed & Fig. 3 (a),(c),(d)\\
   \hline
  FULL-LOW &  1 pure, 1 mixed & Fig. 3 (c),(d)\\
   \hline
  HIGH-HIGH &  1 pure, 2 mixed & Fig. 3 (a),(b),(c)\\
   \hline
  HIGH-MIDDLE & 1 pure, 1 mixed & Fig. 3 (a),(c)\\
   \hline
  MIDDLE-HIGH & 1 pure, 1 mixed & Fig. 3 (a),(b)\\
  \hline
\end{tabular}
\end{table}

\section{Charging Station Pricing Game in Stage I}

Now we analyze the Nash equilibrium of \emph{CSPG} in Stage I, given the NE of {\emph{PSSG}}. Such analysis will lead to the SPE of the entire two-stage game.

To calculate the payoffs of the charging stations, we first derive the demand of each charging station. Take charging station 1 in the FULL-FULL scenario as an example. Let $\Delta p=p_1-p_2$. According to Theorems 1 to 3, the demand of charging station 1 is
\begin{equation}\label{total demand}
 D_1=  \begin{cases}
 2L\lambda d, &\mbox{if $\Delta p \in  [p_{\min}-p_{\max},\theta^L_2]$},\\
 (\omega_1(L-x_2)+L+x_2)\lambda d, &\mbox{if $\Delta p \in  (\theta^L_2,\theta^L_1]$},\\
   (L+x^*)\lambda d, &\mbox{if $\Delta p \in (\theta^L_1,\theta^R_1)$},\\
   (x_1+L)\omega_1\lambda d, &\mbox{if $\Delta p \in  [\theta^R_1,\theta^R_2$)},\\
    0, &\mbox{if $\Delta p \in  [\theta^R_2,p_{\max}-p_{\min}]$}.\\
   \end{cases}
  \end{equation}

Given its competitor's price, a charging station can compute the best price that maximizes its payoffs, defined by (3). We denote the charging station $i$'s best pricing choice as its best response, which is a function of the price $p_j$, \emph{i.e.}, $\mathcal{B}_i(p_j)$. We have
\begin{equation}
\mathcal{B}_i(p_j)\in \arg \max_{p_i\in[p_{\min},p_{\max}]}Q_i(p_i,p_j).
\end{equation}

When the prices of both charging stations are mutual best responses, we have achieved the NE of the CSPG game, denoted by $(p_1^*,p_2^*)$, which satisfies
 \begin{equation}
\mathcal{B}_i(\mathcal{B}_j(p^*_i))=p^*_i, \forall i \in\{1,2\}.
\end{equation}

Next Theorem characterizes the condition for the existence and uniqueness of the pricing equilibrium.\footnote{Even for a convex game, the uniqueness of the NE is not guaranteed. Hence we cannot employ the convexity of $Q_i$ to prove the uniqueness of NE in our game.}
\begin{theorem}
Suppose $i,j \in \{1,2\}$ ($i \neq j$) and consider a region $[a, b]$ with $p_{\min}\leq a < b \leq p_{\max}$. There exists a pure-strategy pricing equilibrium $(p_1^*, p_2^*)$ where $p_1^*\in [a,b]$ and $p_2^*\in [a,b]$, if both of the following conditions are satisfied:

1) $\mathcal{B}_i(\cdot)$ is monotonically increasing in $[a, b]$, for any $i$.

2) $\mathcal{B}_i(\mathcal{B}_j(a))\geq a$ and $\mathcal{B}_i(\mathcal{B}_j(b))\leq b$, for some $i$.

The pure-strategy NE is unique if both conditions 1) and 2) hold and

3)  $\mathcal{B}_i(p_j)-p_j$ is strictly monotonically decreasing in $[a, b]$, for any $i$.\footnote{$\mathcal{B}_i(p_j)$ does not need to be a strictly monotonic function.}
\end{theorem}

The proof of Theorem 6 can be found in Appendix F. In most of our simulations, conditions 2) and 3) are satisfied by $a=p_{\min}$ and $b=p_{\max}$. We can view $\mathcal{B}_i(p_j)-p_j$ as the \emph{best price offset} of charging station $i$. Condition 3) implies that when the competitor's price increases, a charging station's best price offset will decrease. This means that although charging station $i$ might increase its price $p_i$ by responding to its competitor station $j$'s price increase, the price gap will reduce. Condition 1), however, is not always easy to satisfy in simulations. However, we note that conditions 1) -3) are sufficient but not necessary conditions, and a pricing equilibrium may exist even if these conditions are not satisfied (as observed from our simulations). For the simplicity of analysis, we will assume that all three conditions in Theorem 6 are satisfied by $a=p_{\min}$ and $b=p_{\max}$.

\section{Computing the Equilibrium}
As characterizing the closed-form pricing equilibrium in CSPG in Stage I is very challenging, next we will focus on developing a low-complexity algorithm to compute the pricing equilibrium.
\newtheorem{proposition}{\textbf{Proposition}}
\begin{proposition}
Let $\Theta_i(p_i)=\mathcal{B}_i(\mathcal{B}_j(p_i))-p_i$. For any price $p_i$ in $[p_{\min}, p_{\max}]$, we have $\Theta_i(p_i)<0$ if $p_i>p^*_i$, and $\Theta_i(p_i)>0$ if $p_i<p_i^\ast$.
\end{proposition}

The proof of Proposition 1 is given in Appendix G. Based on Proposition 1, a charging station $i$ can figure out whether its current price $p_i$ is larger or smaller than the equilibrium price, by evaluating the function of $\Theta_i$.

With Proposition 1, we propose an iterative Directional SPE Search Algorithm (DSSA) that searches $p_1^\ast$ in the region of $[p_{\min}, p_{\max}]$. In the $t$th iteration of DSSA, the algorithm updates $p_1(t)$ by
\begin{equation}\label{iteration}
p_1(t+1)=[p_1(t)+d(t)\delta(t)]^{p_{\max}}_{p_{\min}},
\end{equation}
where
$[x]^a_b= \min(\max(x,b),a)$,
\begin{equation}\label{update_delta}
\delta(t)=
   \begin{cases}
   \delta(t-1), & \mbox{if $\Theta_1(p_1(t)) \cdot \Theta_1(p_1(t-1))> 0$},\\
   \alpha \delta(t-1), & \mbox{if $\Theta_1(p_1(t)) \cdot \Theta_1(p_1(t-1))<0 $}, \\
   \end{cases}
\end{equation}
and
\begin{equation}\label{update_d}
d (t)=
   \begin{cases}
   1, & \mbox{if $\Theta(p_1(t))> 0$},\\
   -1, & \mbox{if $\Theta(p_1(t))< 0$},\\
   0, & \mbox{if $\Theta(p_1(t))= 0$}.
   \end{cases}
\end{equation}
In the above equations, $d(t)$ represents the search direction, $\delta(t)$ is the step size and $\alpha$ is a constant in $(0,1)$. $\Theta_1(p_1(t)) \cdot \Theta_1(p_1(t-1))<0 $ indicates that DSSA has leaped over $p^*_1$ in the previous iteration. Accordingly, DSSA changes its search direction and continues to search $p_1^\ast$ with a smaller step size.

 We show the details of DSSA in Algorithm 1. Notice that each charging station $i$ executes Algorithm 1 independently without synchronization. In addition, if the charging stations' locations, service capacity, unit electricity cost, and feasible price range are public information, DSSA does not require any explicit information exchange between the charging stations, as each charging station $i$ can compute the function $\Theta_i$ locally.

\begin{algorithm}
\caption{DSSA for Station $i\in \{1,2\}$}
 \KwIn{$L, x_i, \mu_i, k_i, c_i, \lambda, p_{\min}, p_{\max},\alpha,\delta(0),\varepsilon$}
 \KwOut{$p_1^*,p_2^*$ }
 \If {$|\Theta_i(p_{\min})|\leq \varepsilon$ } {
    $p_1^*=p_2^*=p_{\min}$ and terminate\;
 }
 \If {$|\Theta_i(p_{\max})|\leq \varepsilon$} {
    $p_1^*=p_2^*=p_{\max}$ and terminate\;
 }
 Set $\Theta_i(p_i(0))=1$, $t=1$ and randomly choose $p_i(1)$ from $(p_{\min},p_{\max})$\;

 \While{$|\Theta_i(p_i(t))|/p_i(t)> \varepsilon$}{
    Update $d(t)$ and $\delta(t)$ with \eqref{update_d} and \eqref{update_delta}, respectively\;
    Update $p_i(t+1)$ with \eqref{iteration}, and $t=t+1$\;
 }
 $p_i^*=p_i(t)$ and $p_j^*=\mathcal{B}_j(p_i^\ast)$;
\end{algorithm}

\begin{theorem}
The DSSA algorithm converges to an SPE of our Stackelberg game Q-linearly.
\end{theorem}

The proof of Theorem 7 is given in Appendix H.

\section{Numerical Results}

We numerically verify our equilibrium analysis and the performance of the proposed algorithm. Unless specified otherwise, we choose system parameters as follows: $L=10$, $x_1=-8$, $x_2=5$, $k_1=k_2=2$, $d=60$~\footnote{For EVs produced by Tesla, the battery capacities can be 60 kwh, 70 kwh, 85 kwh, etc. Here we assume $d$ to be 60 kwh.}, $k_p=4$, $k_q=5$, $k_l=1.5$, $\lambda=1$, $c_1=c_2=0.15$, $p_{min}\geq 0.15$, $p_{max}\leq 0.3$~\footnote{In some Chinese cities, the electricity price is about 0.15 US dollar, and the maximally allowed charging price is about 0.3 US dollar.}, $\check{c}_1=\check{c}_2=1$, and $\varepsilon=0.001$.

\subsection{Charging Station Selection Equilibrium in Stage II}

\subsubsection{FULL-FULL}
We first consider the FULL-FULL scenario with $\mu_1=16$ and $\mu_2=14$, in which both charging stations are able to serve all PEVs. Fig. 4 and Fig. 5 show the indifference point location $x^*$ and the probability of selecting charging station 1 ($\omega_1$) under various values of price difference $p_1-p_2$, respectively. In this simulation, $\theta_2^L=-0.1063$, $\theta_1^L=-0.0981$, $\theta_1^R=0.0881$, and $\theta_2^R=0.0988$.

Fig. 4 shows when $p_1-p_2$ increases from $\theta_1^L$ to $\theta_1^R$, $x^{\ast}$ decreases from $x_2$ to $x_1$. This corresponds to the result in Theorem 1, and shows that a larger price difference $p_1-p_2$ will lead to more PEVs choosing charging station 2.

Fig. 5 provides the values of probability $\omega_1$ of the PEVs in the ranges of $[-L, x_1]$ and $[x_2, L]$, when $p_1-p_2$ is either smaller than $\theta_1^L$ or larger than $\theta_1^R$. More specifically, when $p_1-p_2\in (\theta_2^L,\theta_1^L)$, $\omega_1$ of the PEVs in $[x_2, L]$ is between 1 and 0. In other words, all the PEVs in $[-L, x_2]$ select station 1, and the PEVs in $[x_2, L]$ select station 1 with probability $\omega_1$ and station 2 with probability $1-\omega_1$. When $p_1-p_2\in (\theta_1^R,\theta_2^R)$, $\omega_1$ of the PEVs in $[-L, x_1]$ is between 1 and 0. These results are consistent with Theorem 2. When $p_1-p_2\geq \theta_2^R$ or $p_1-p_2\leq \theta_2^L$, we will have a pure-strategy equilibrium in which all PEVs choose station 2 or station 1.

\begin{figure}
\centering
\scalebox{1.4}[1.4]{\includegraphics[width=1.8in]{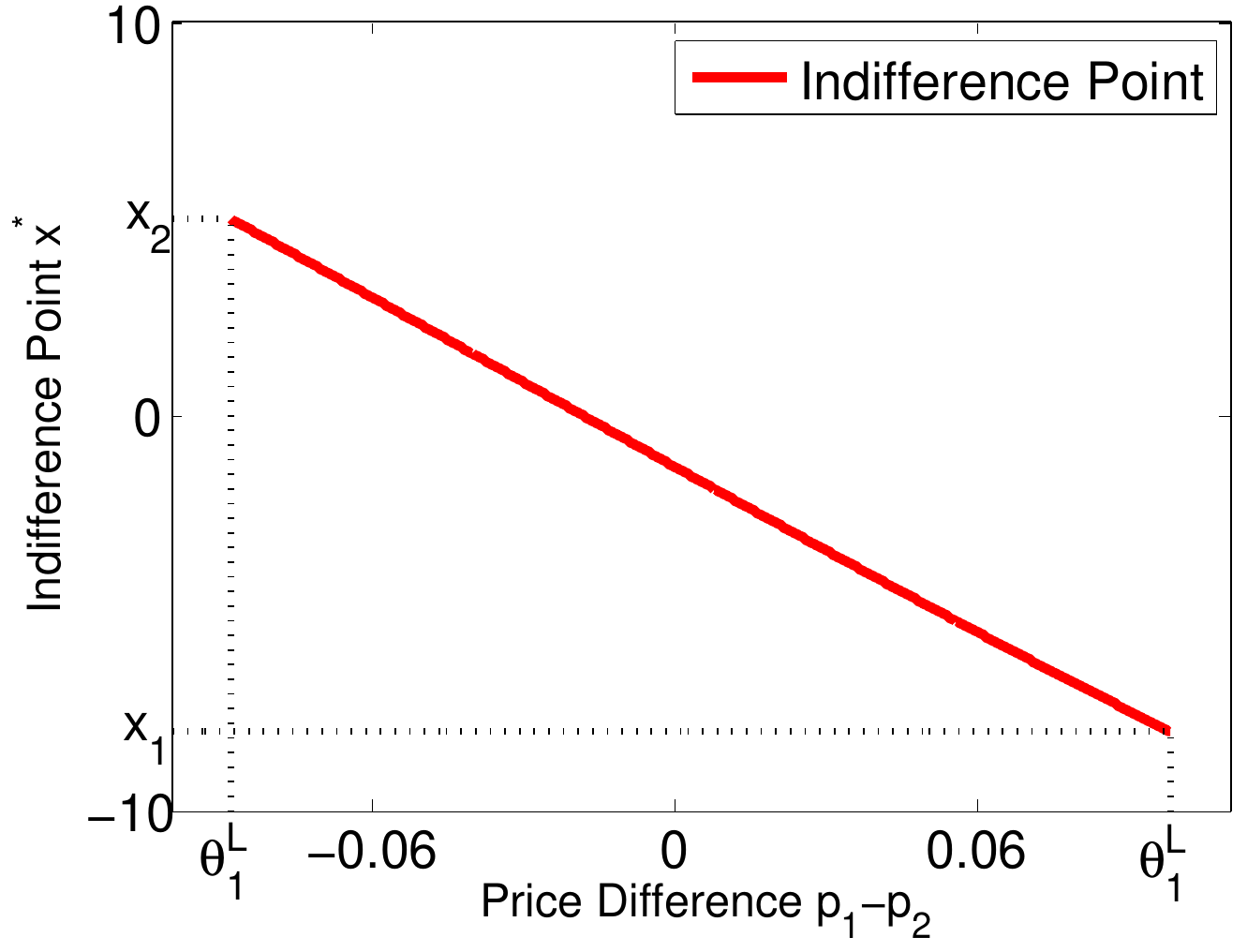}}
\caption{ Indifference point $x^*$ in the FULL-FULL scenario.}
\end{figure}%

\begin{figure}
\centering
\scalebox{1.4}[1.4]{\includegraphics[width=1.8in]{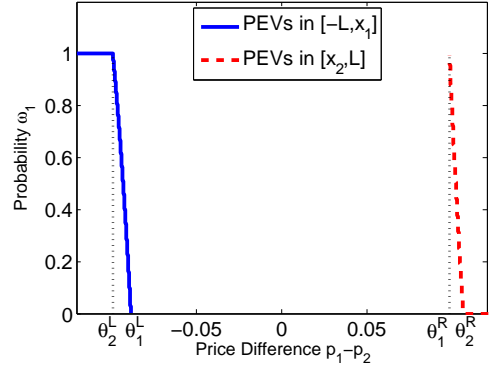}}
\caption{Probability $\omega_1$ in the FULL-FULL scenario.}
\end{figure}%

\begin{figure}
\centering
\scalebox{1.4}[1.4]{\includegraphics[width=1.8in]{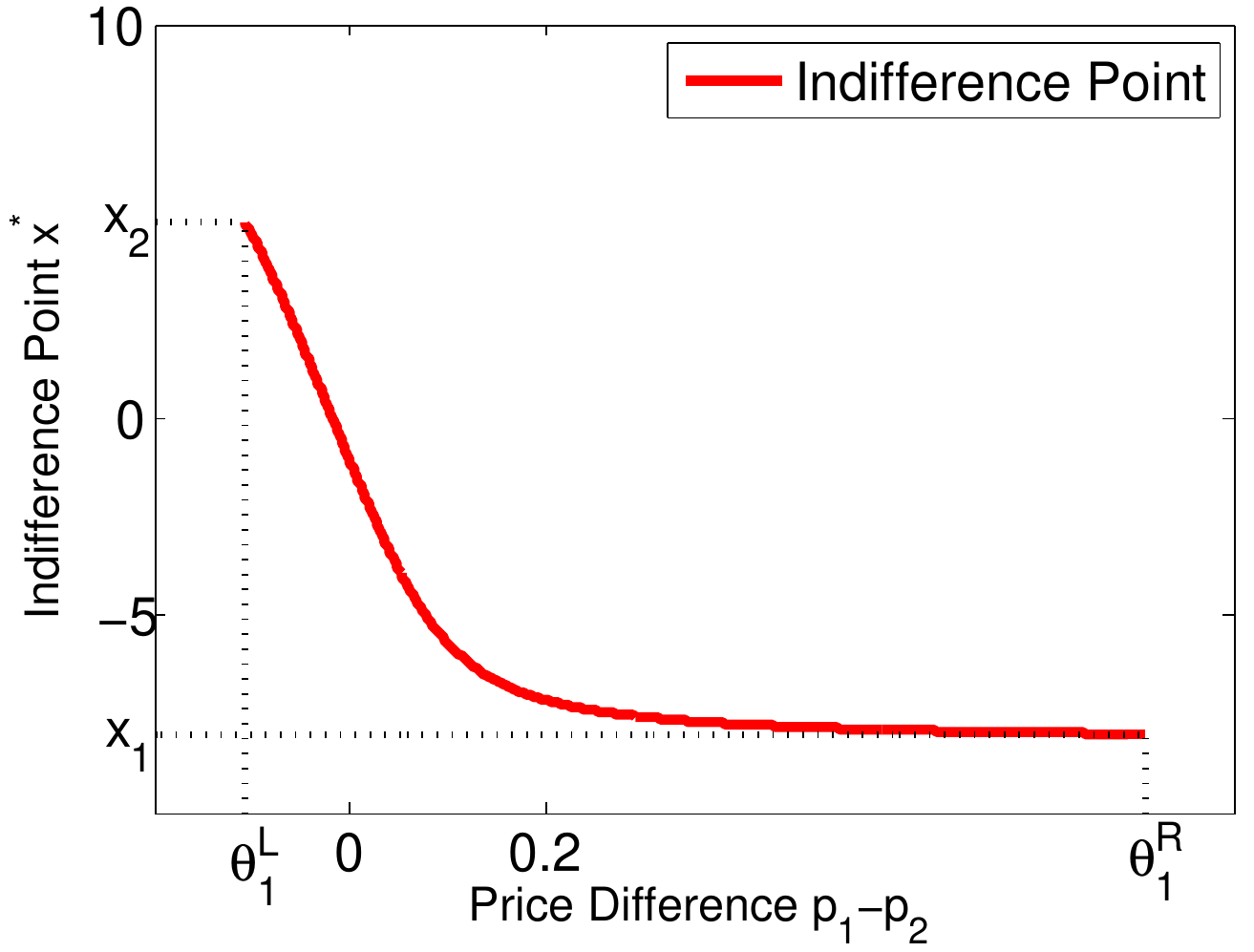}}
\caption{ Indifference point $x^*$ in the HIGH-HIGH scenario.}
\end{figure}%

\begin{figure}
\centering
\scalebox{1.4}[1.4]{\includegraphics[width=1.8in]{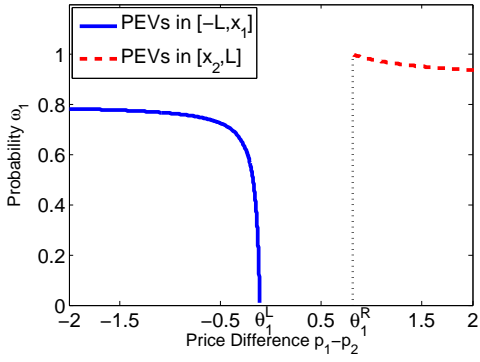}}
\caption{Probability $\omega_1$ in the HIGH-HIGH scenario.}
\end{figure}

\subsubsection{Other Scenarios}
Now we consider the HIGH-HIGH scenario with $\mu_1=9.5$ and $\mu_2=9.1$. Similar to Fig. 4, Fig. 6 illustrates how the indifference point $x^{\ast}$ changes with $p_1-p_2$ in this scenario. It can be seen that the indifference point is in the range of $[x_1,x_2]=[-8,5]$. In Fig. 7, we provide the values of $\omega_1$ of the PEVs in the ranges of $[-L, x_1]$ and $[x_2, L]$. Different from Fig. 5, however, Fig. 7 indicates that $\omega_1$ cannot continuously span the entire range of $[0,1]$ when $p_1-p_2$ changes, since neither of the stations is able to serve all the PEVs. In fact, when $p_1-p_2<\theta_1^L$, we always have $\omega_1<0.8=\frac{k_1\mu_1-(L+x_2)\lambda}{(L-x_2)\lambda}$. When $p_1-p_2>\theta_1^R$, we always have $\omega_1>0.9=\frac{2L\lambda-k_2\mu_2}{(L+x_1)\lambda}$. This is consistent with Theorem 4.

Fig. 8 shows the location of the indifference point in the MIDDLE-MIDDLE scenario with $\mu_1=7$ and $\mu_2=6$. It is obvious that the indifference point is in the range of $(-2,4)$, which coincides with $(L-\frac{k_2\mu_2}{\lambda}, \frac{k_1\mu_1}{\lambda}-L)$. This is consistent with Theorem 5.

Finally, we consider the HIGH-LOW scenario with $\mu_1=9$ and $\mu_2=2$. Fig. 9 provides the values of $\omega_1$ of the PEVs in the range of $[x_2, L]$ in this scenario. It can be seen that $\omega_1$ is within the range of $(0.2, 0.6)$, which corresponds to $(1-\frac{k_2\mu_2}{L-x_2}, \frac{k_1\mu_1-(L+x_2)\lambda}{L-x_2})$. Fig. 9 tells that if a station has a sufficient service capacity but its competitor is limited in service capacity, it can always attract some PEVs no matter how high its price is. In our simulation, station 2 has a much lower service capacity. Therefore, the PEVs on the right side of station 2 still choose station 1 with a probability greater than $0.2$ even when the price difference is relatively high, e.g. $3$.

\begin{figure}
\centering
\scalebox{1.4}[1.4]{\includegraphics[width=1.8in]{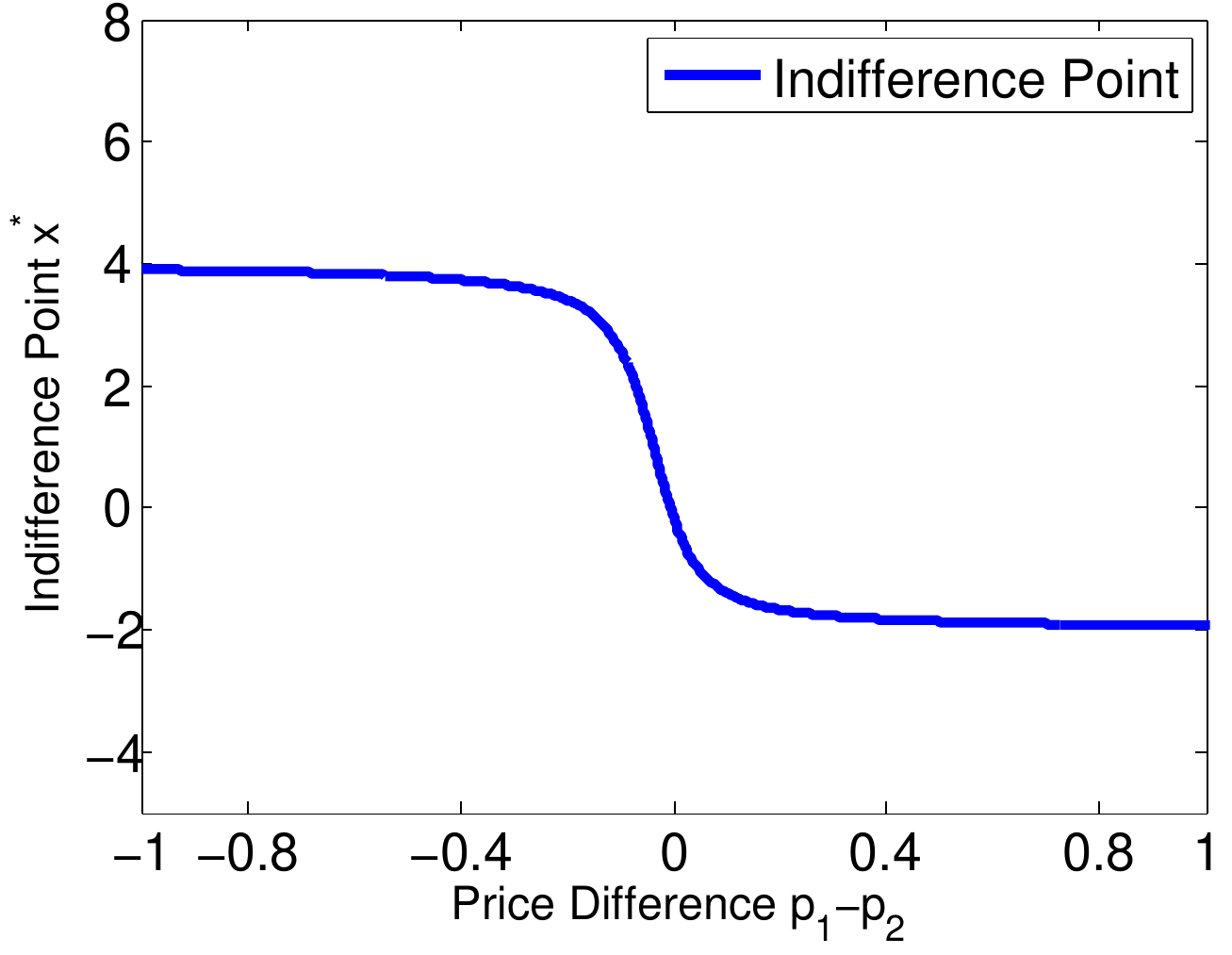}}
\caption{ Indifference point $x^*$ in the MID-MID scenario.}
\end{figure}

\begin{figure}
\centering
\scalebox{1.4}[1.4]{\includegraphics[width=1.8in]{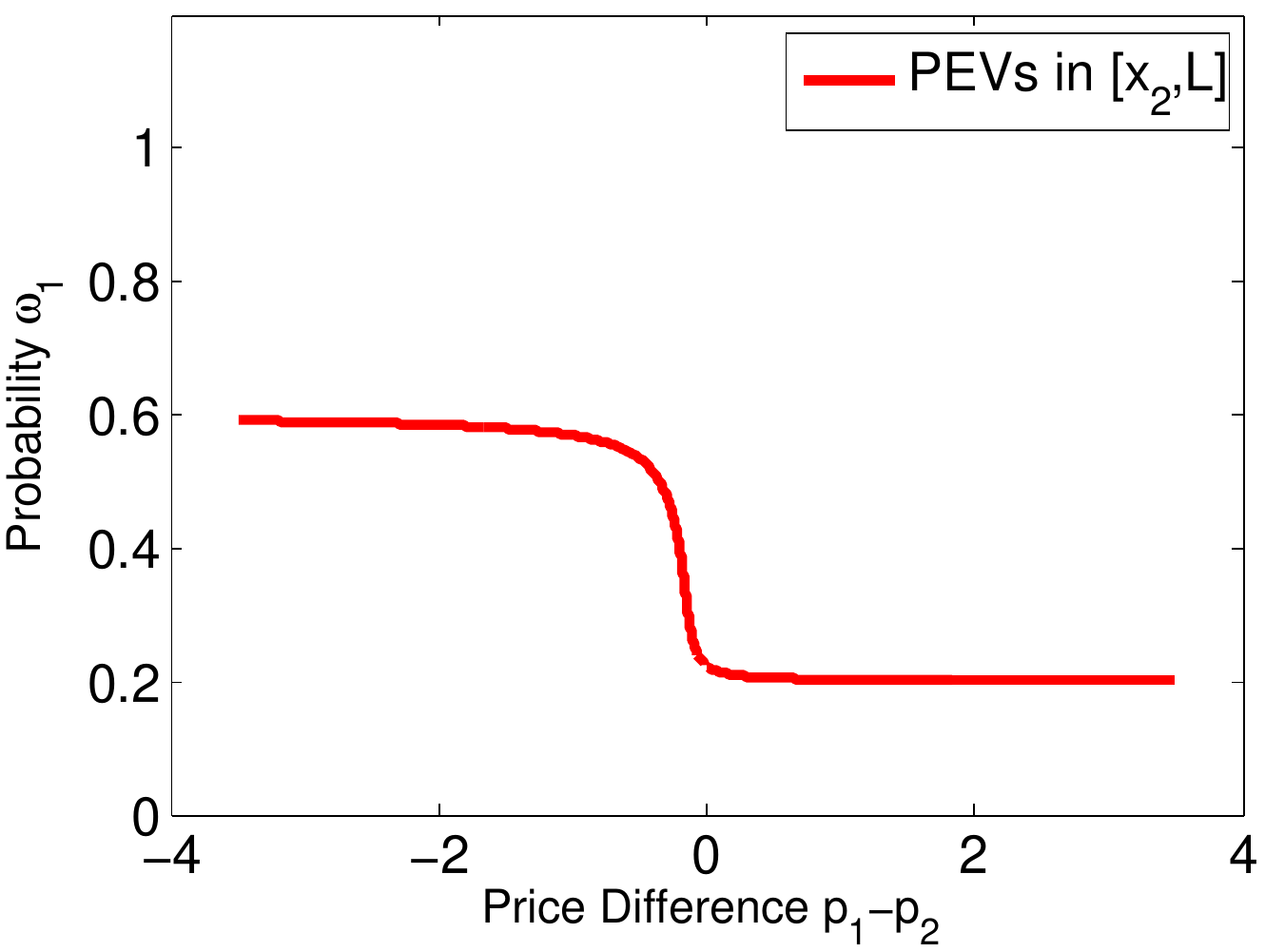}}
\caption{Probability $\omega_1$ in te HIGH-LOW scenario.}
\end{figure}

\subsection{Pricing Equilibrium of CSPG}

\begin{figure}
\centering
\scalebox{1.4}[1.4]{\includegraphics[width=1.8in]{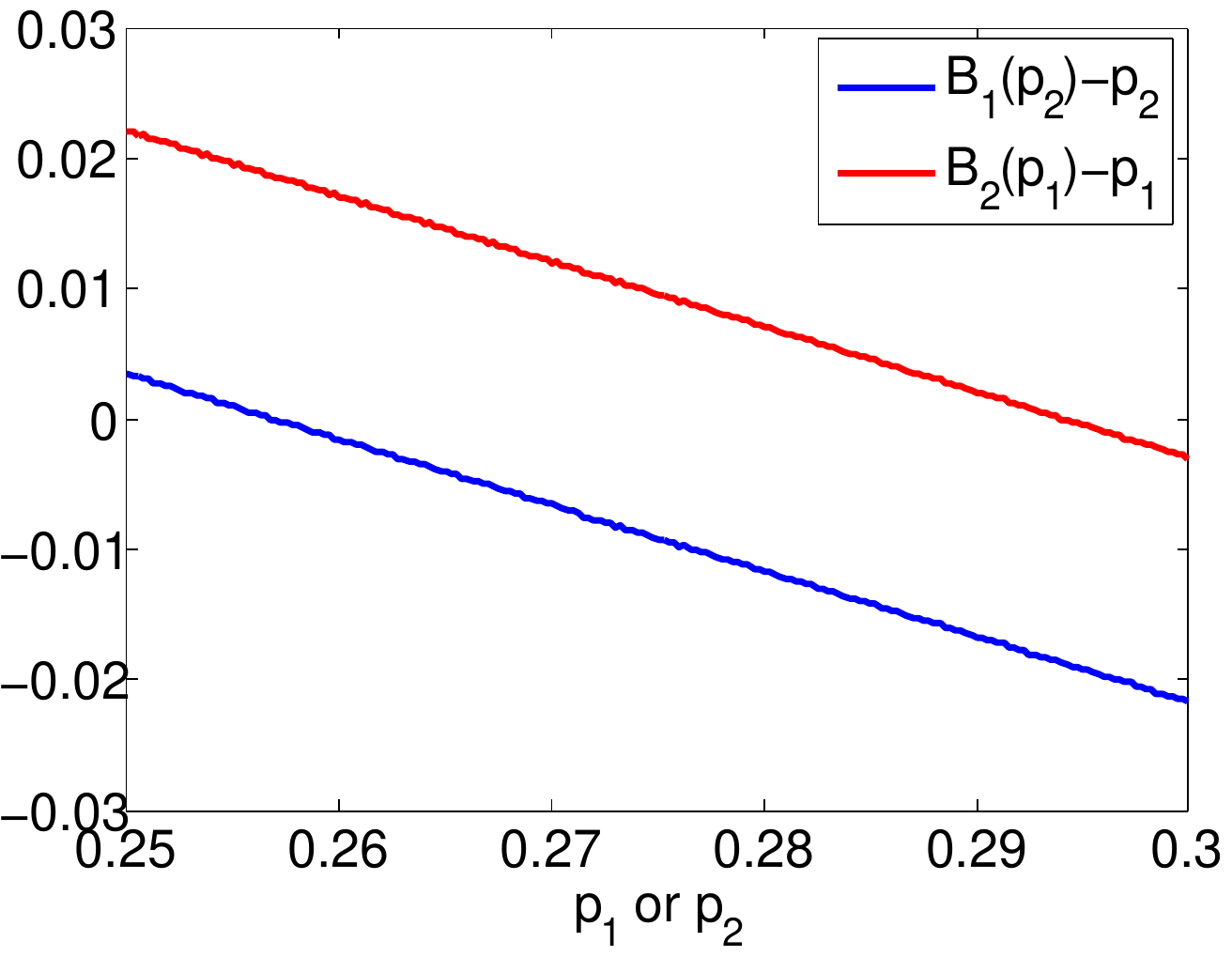}}
\caption{Best price offsets with $\mu_1=16$, $\mu_2=14$ and $p_i \in [0.25, 0.3]$.}
\end{figure}

\begin{figure}
\centering
\scalebox{1.4}[1.4]{\includegraphics[width=1.8in]{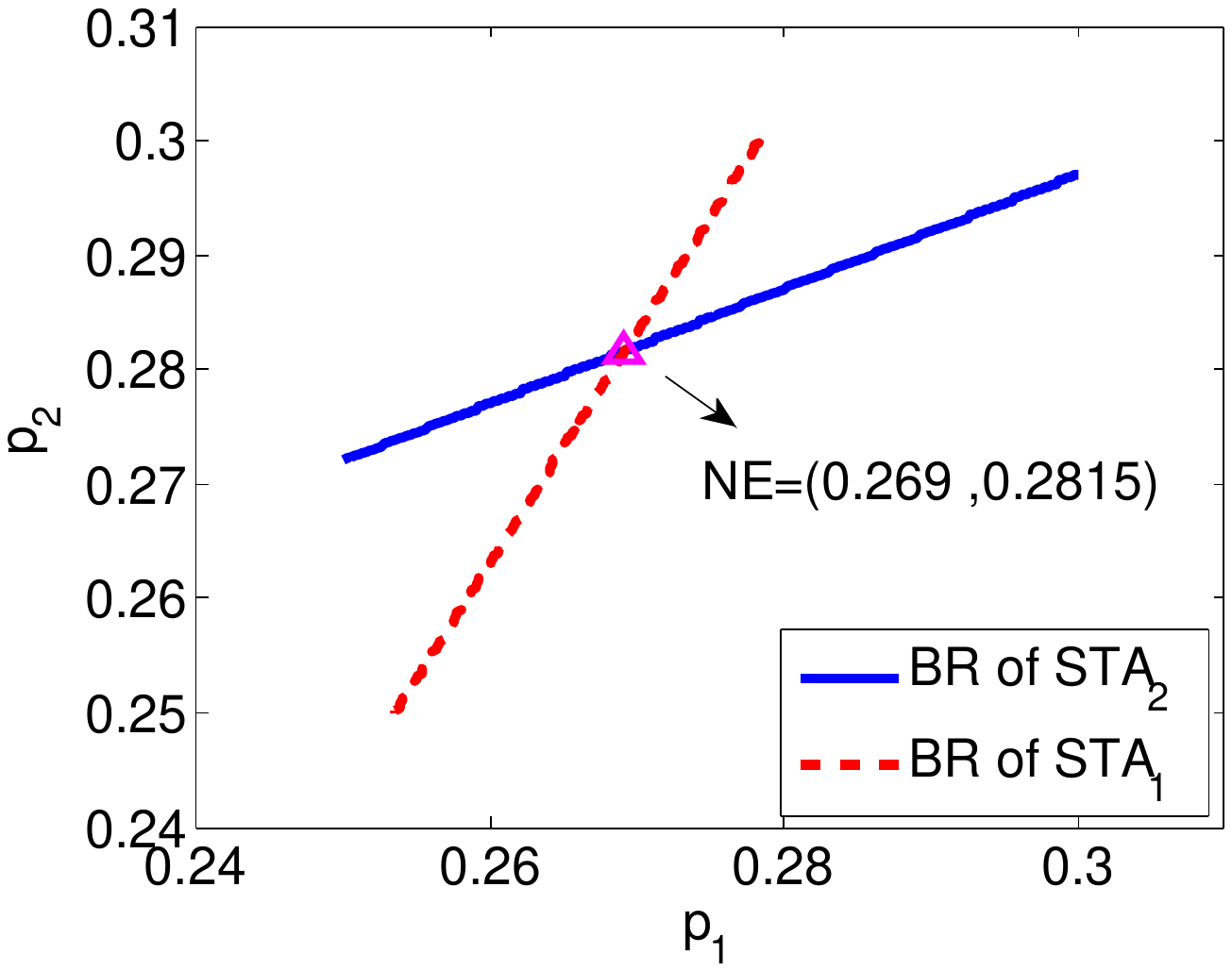}}
\caption{Best response curves with $x_1=-8$, $x_2=5$ and $p_i \in [0.25, 0.3]$.}
\end{figure}

\begin{figure}
\centering
\scalebox{1.4}[1.4]{\includegraphics[width=1.8in]{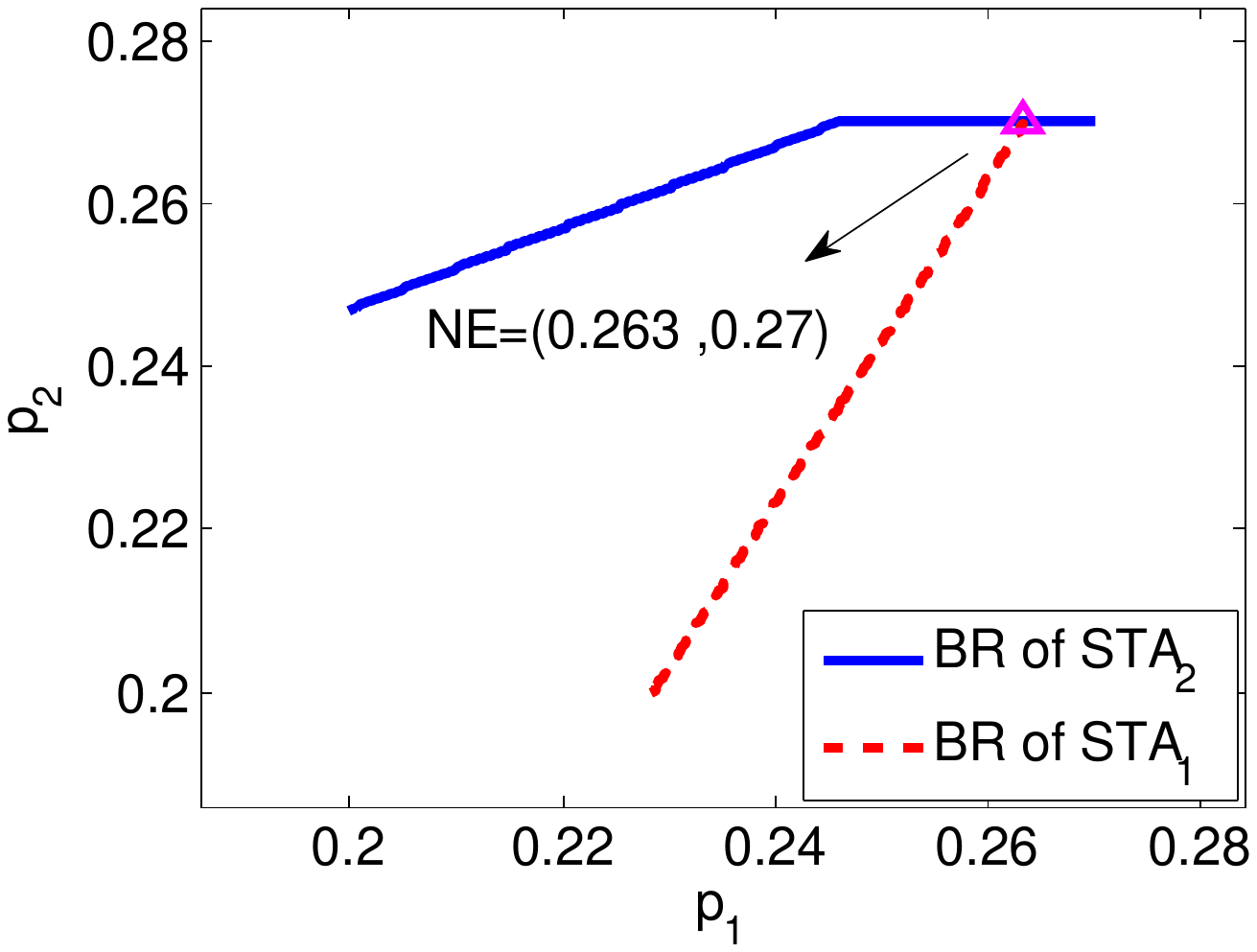}}
\caption{Best response curves with $x_1=-8$, $x_2=5$ and $p_i \in [0.2, 0.27]$.}
\end{figure}

\begin{figure}
\centering
\scalebox{1.4}[1.4]{\includegraphics[width=1.8in]{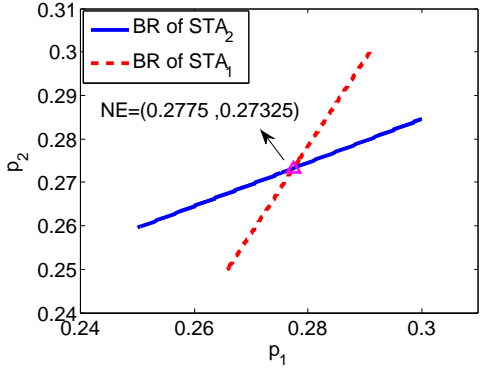}}
\caption{ Best response curves with $x_1=-8$, $x_2=9$ and $p_i \in [0.25, 0.3]$.}
\end{figure}

We first illustrate the conditions in Theorem 6. Let us consider the FULL-FULL scenario with $\mu_1=16, \mu_2=14$, and $p_i\in[0.25, 0.3]$. Fig. 10 and Fig. 11 show the best price offset and the best response, respectively. According to Fig. 10, the best price offset, \emph{i.e.}, $\mathcal{B}_i(p_j))-p_j$ ($i\neq j$), is strictly decreasing with its competitor's price. As shown in Fig. 11, the best response curves of charging station 1 and 2 are both increasing.

Next we numerically compute the NE of \emph{CSPG} through computing the intersection of the best response functions of both charging stations. In Fig. 11, $(p_{min},p_{max})=(0.25,0.3)$, and the intersection of the best response curves corresponds to the NE $(0.269, 0.282)$. Now we consider the other two FULL-FULL scenarios.
\begin{itemize}
\item Fig. 12 shows the best response curves in the scenarios with $\mu_1=16$, $\mu_2=14$ and $(p_{min},p_{max})=(0.2,0.27)$. It can be seen that the intersection of the best response curves indicates the NE $(0.26, 0.27)$. In this simulation, $p_{max}$ is small and the candidate prices are relatively low. As a result, the best price offset is often positive. This means that each charging station has an incentive to further increase its price. This explains why a high equilibrium price is adopted by each station.
\item In Fig. 13, we characterize the best response curves in the scenario with $\mu_1=16$, $\mu_2=14$ and $(p_{min},p_{max})=(0.25,0.3)$. Different from the case in Fig. 11, we set the location of station 2 to $x_1=9$ instead of $x_1=5$ in Fig. 13. Now we compare the NEs shown in Fig. 11 and Fig. 13. Although station 2 declares a higher equilibrium price than its competitor in Fig. 11, the opposite outcome is true in Fig. 13. This is because station 2 is relatively far away from the middle point in the case of Fig. 13 than the case of Fig. 11, hence needs to announce a smaller price to attract the PEVs.
\end{itemize}

\subsection{The DSSA Algorithm}

\begin{figure}
\centering
\scalebox{1.4}[1.4]{\includegraphics[width=1.8in]{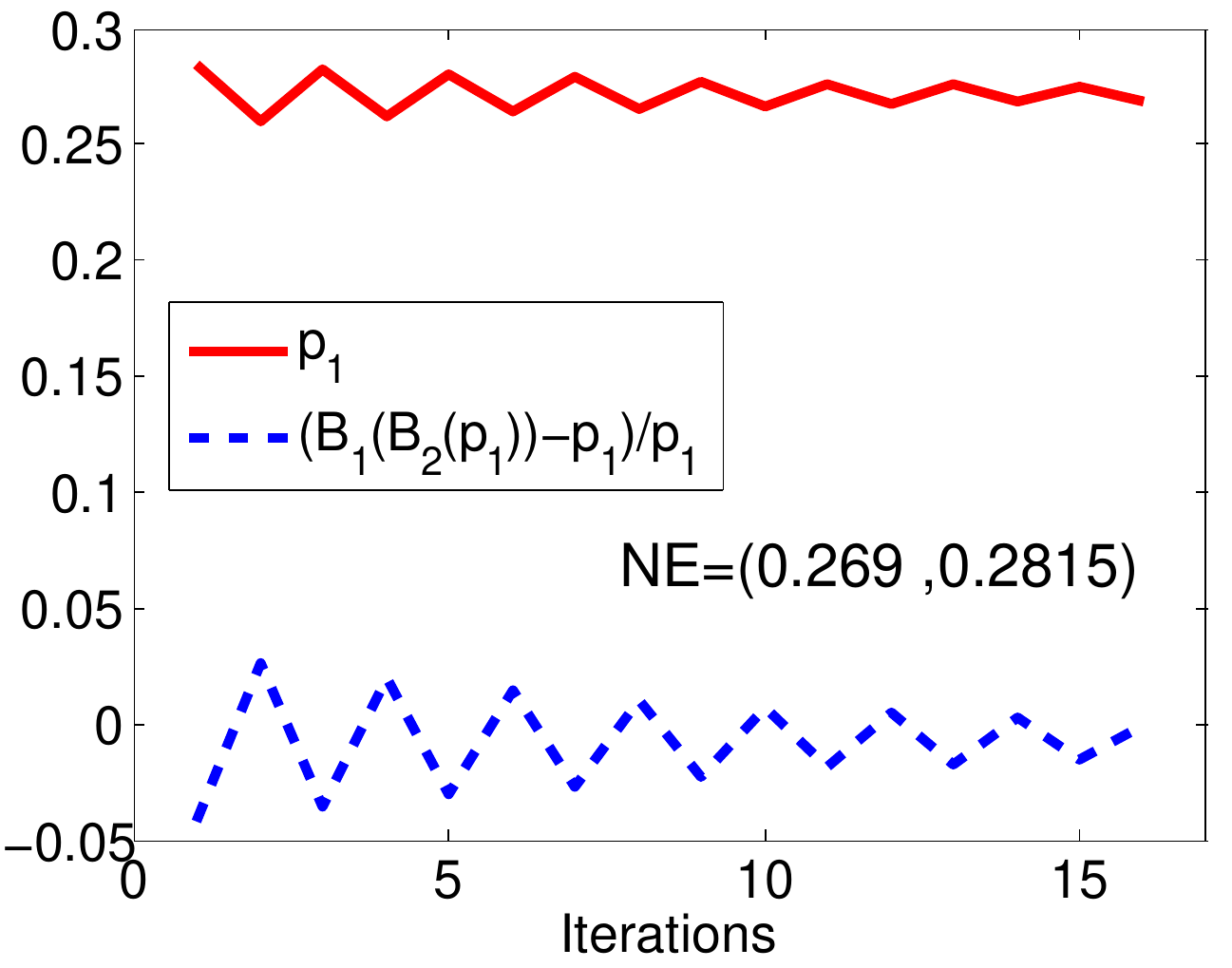}}
\caption{Iterations of DSSA with $x_1=-8$, $x_2=5$ and $p_i\in[0.25,0.3]$.}
\end{figure}

\begin{figure}
\centering
\scalebox{1.4}[1.4]{\includegraphics[width=1.8in]{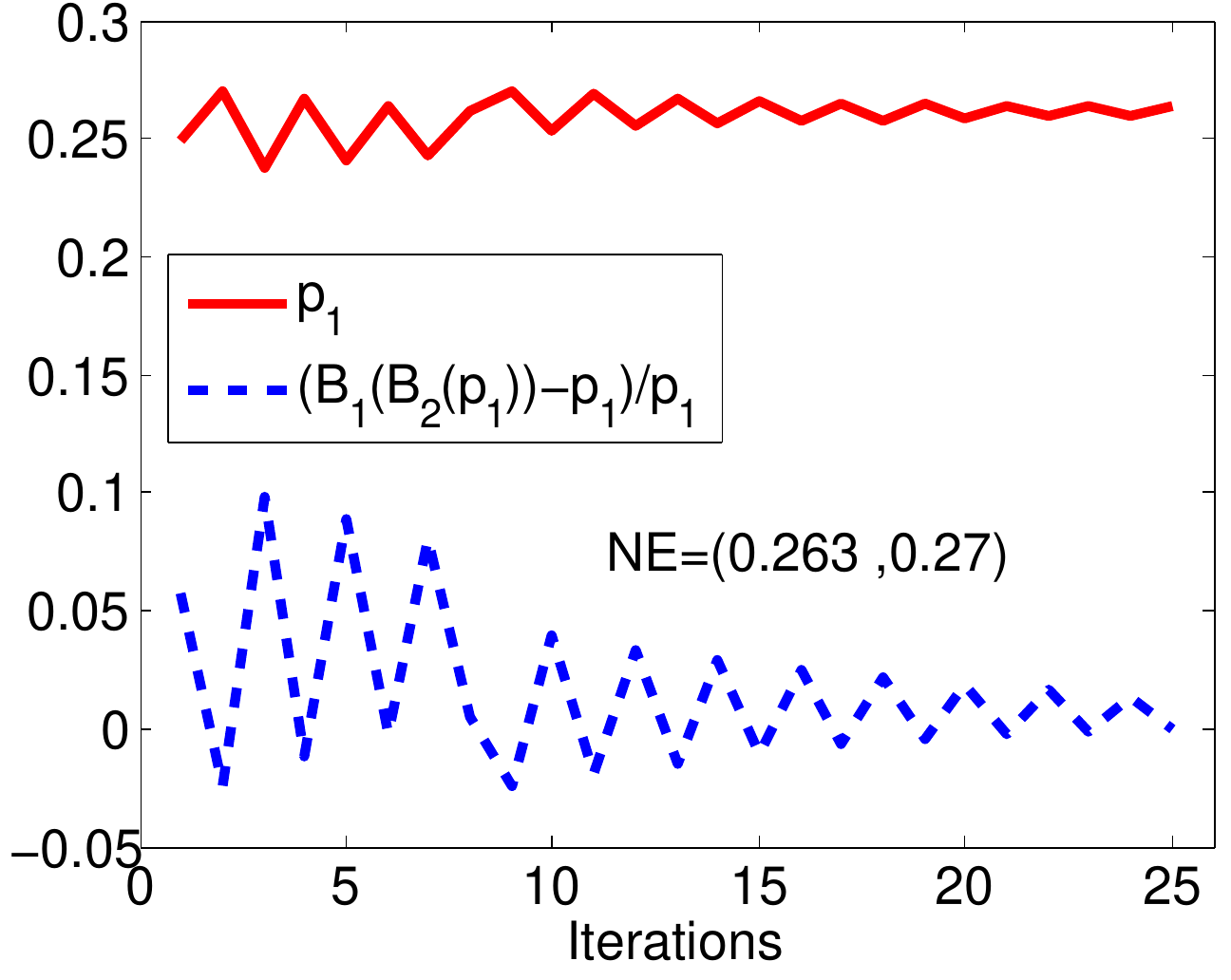}}
\caption{Iterations of DSSA with $x_1=-8$, $x_2=5$ and $p_i\in[0.2,0.27]$.}
\end{figure}

\begin{figure}
\centering
\scalebox{1.4}[1.4]{\includegraphics[width=1.8in]{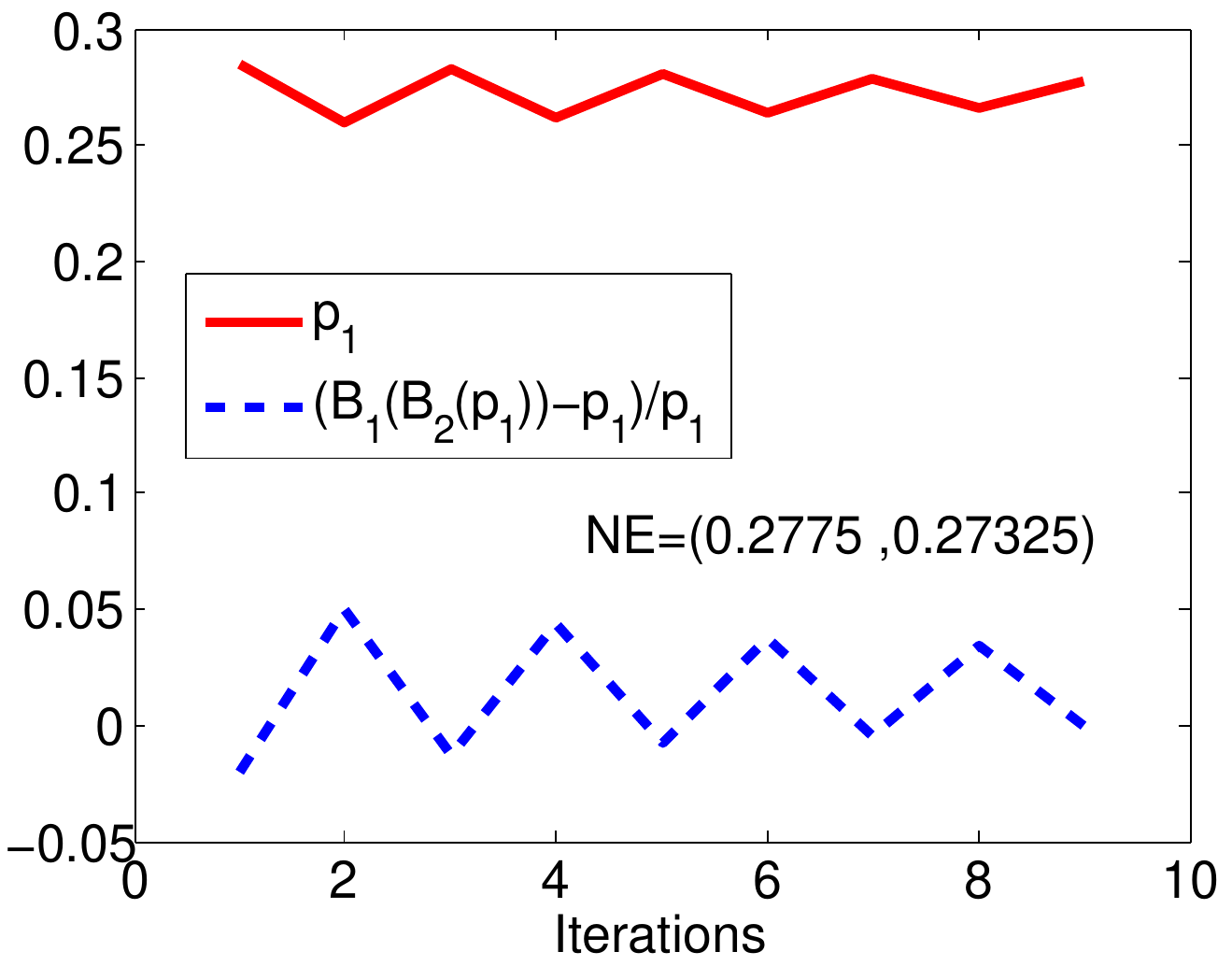}}
\caption{Iterations of DSSA with $x_1=-8$, $x_2=9$ and $p_i\in[0.25,0.3]$.}
\end{figure}

Finally, we demonstrate the convergence and computational efficiency of DSSA. As mentioned in Section VI, when $|\mathcal{B}_i(\mathcal{B}_j(p_i))-p_i|/p_i\leq \varepsilon$, DSSA converges to the NE. We consider the scenarios corresponding to Fig. 11 - Fig. 13, and show the iterations of DSSA in Fig. 14 - Fig. 16, respectively. In all cases, DSSA converges to the NEs within 25 iterations. Furthermore, $\mathcal{B}_1(\mathcal{B}_2(p_1))-p_1$ is negative when $p_1>p_1^*$, and is positive when $p_1<p_1^*$. When $\mathcal{B}_1(\mathcal{B}_2(p_1))-p_1$ approaches zero, DSSA terminates at $p_1^*$. The above phenomena verify the principle of our DSSA algorithm.

\section{Conclusion}

This work studies the charging station pricing and PEV station selection through a two-stage Stackelberg game. We characterize the charging station selection equilibrium in Stage II, and characterize the sufficient conditions for the existence of the pricing equilibrium in Stage I. We also develop a low complexity algorithm to compute the equilibria of the entire game. In the future work, we will focus on more general system models considering more than two charging stations, networks of roads, and the impact of traffic conditions.

\begin{appendices}
\section{Proof of Theorem 1}
\newtheorem{remark}{\textbf{Remark}}
\begin{remark}
Suppose a PEV $n$ at $x$ adopts a strategy $s_n^*$ at an NE in \emph{PSSG}. If $s_n^*=1$, all the PEVs at $x^{'}<x$ take the strategy $s_n^*(x^{'})=1$ at that NE. If $s_n^*=2$, all the PEVs at $x^{'}>x$ take the strategy $s_n^*(x^{'})=2$ at that NE.
\end{remark}

According to Remark 1, we know that if when an indifference point exists and a corresponding NE is achieved, the PEVs on the left side of the indifference point choose station 1 and the PEVs on the right side of the indifference point choose station 2.

Next we will first prove that the root to \eqref{calculate_indifference} is unique. Then we will show that the strategy profile corresponding to (4) is an NE. Finally, we will prove that no other NE exists.

\subsection{Existence and uniqueness of the root to \eqref{calculate_indifference}}
For convenience, let $f(x)=k_p d (p_1-p_2)+k_q(q_1(x+L)-q_2(L-x))+k_l(2x-x_1-x_2)$. We have
\begin{equation}
\begin{aligned}
f(x_1)&=k_p d (p_1-p_2)+k_q(q_1(x_1+L)-q_2(L-x_1))\\
&+k_l(x_1-x_2)\\
&<k_p d \theta_1^R+k_q(q_1(x_1+L)-q_2(L-x_1))\\
&+k_l(x_1-x_2)=0\\
\end{aligned}
\end{equation}
Similarly, we can prove that $f(x_2)>0$. Hence we can conclude that $f(x)=0$ holds for some $x \in (x_1,x_2)$.

Furthermore, $f(x)$ is a strictly monotonic increasing function. Hence there exists one unique root satisfying $f(x) = 0$.

\subsection{Nash equilibrium}
If every PEV adopts the strategy described by (4), $|A_1|=x^*+L$ and $|A_2|=L-x^*$. We first consider the PEVs on the left side of the indifference point. If a PEV $n$ at $x$ selects station 1, its payoff is $U_n(1,s_{-n})=-k_p d p_1-k_q q_1(|A_1|) -k_l |x-x_1|$. If it selects station 2, its payoff is $U_n(2,s_{-n})=-k_p d p_2-k_q q_1(|A_2|) -k_l |x-x_2|$.
\begin{enumerate}
\item When $x\geq x_1$, we have $U_n(2,s_{-n})-U_n(1,s_{-n})=k_p d (p_1-p_2)+k_q(q_1(|A_1|)-q_2(|A_2|))+k_l(2x-x_2-x_1)<k_p d (p_1-p_2)+k_q(q_1(|A_1|)-q_2(|A_2|))+k_l(2x^*-x_1-x_2)=0$.
\item When $x<x_1$, we have $U_n(2,s_{-n})-U_n(1,s_{-n})=k_p d (p_1-p_2)+k_q(q_1(|A_1|)-q_2(|A_2|))+k_l(x_1-x_2)<k_p d (p_1-p_2)+k_q(q_1(|A_1|)-q_2(|A_2|))+k_l(2x^*-x_1-x_2)=0$.
\end{enumerate}
Hence, PEV $n$ prefers station 1 to station 2.

Similarly, we can prove that the PEVs on the right side of the indifference point prefers station 2 to station 1. In all, with the strategy described by (4), no PEV has an incentive to unilaterally change its station selection. That is, the strategy profile corresponding to (4) is an NE.

\subsection{Uniqueness of Nash equilibrium}
 We have proved that the NE is unique when an indifference point exists. Now we consider the case where indifference point does not exist.

The indifference point exists in the scenario where the PEVs in the range of $[-L, x^*)$ or $(x^*, L]$ choose the same equilibrium strategy. The situation where no indifference point exists will emerge when some PEVs take different equilibrium strategy. In the following, we prove that this situation will never emerge by contradiction.

We first consider the case where a different pure-strategy NE exists. Suppose at least one PEV at $x<x^*$ selects station 2 at such NE. Assume that among these PEVs, PEV $n$ has the smallest position, \emph{i.e.}, $\tilde{x}$. Clearly, all the PEVs on the right side of PEV $n$ select station 2 at this NE since they are closer to station 2 than PEV $n$. As for the PEVs on the left side of PEV $n$, all of them select station 1~\footnote{Otherwise, PEV $n$ is not the one with the smallest position.}. This results in the situation where the PEVs in the segment $[-L, \tilde{x})$ or $(\tilde{x}, L]$ choose the same equilibrium strategy, which is impossible. Similarly, we can analyze the case where at least one PEV at $x>x^*$ selects station 1 when a different NE is achieved.

The proof for the case where some PEV chooses a mixed strategy at a different NE is similar and hence is omitted.

\section{Proof of Theorem 2}

We consider the case of $p_1-p_2 \in [\theta^R_1,\theta^R_2)$. The proof in the case of $p_1-p_2 \in (\theta^L_2,\theta^L_1]$ is similar, and is omitted here.
\subsection{Existence and uniqueness of the root to \eqref{probability_indifference}}
For convenience, let $f(\omega)=k_pd(p_1-p_2)+k_l(x_1-x_2)+k_q\cdot(q_1((x_1+L)\omega)-q_2(L-x_1+(x_1+L)(1-\omega)))$.
We have
\begin{equation}
\begin{aligned}
f(0)&=k_p d(p_1-p_2)+k_l(x_1-x_2)-k_q\cdot q_2(L-x_1+(x_1+L))\\
&<k_p d \theta_2^R -k_l(x_2-x_1)-k_q\cdot q_2(L-x_1+(x_1+L))=0
\end{aligned}
\end{equation}
Similarly, we have $f(1)\geq 0$. Hence we can conclude that there exists an $\omega \in [0,1]$ satisfying $f(\omega)=0$. Furthermore, $f(\omega)$ is a strictly monotonic increasing function. Hence there exists one unique $\omega \in [0,1]$ satisfying $f(\omega)=0$.
In addition, it can be proved that $\omega \in [0,1]$ satisfying $f(\omega)=0$ cannot be less than 0 or greater than 1.

\subsection{Nash equilibrium}

When every PEV adopts the strategy given by (6), $|A_1|=\omega_1(x_1+L)$ and $|A_2|=(1-\omega_1)(x_1+L)+L-x_1$. In the following, we consider a PEV $n$ at $x$ in two cases: 1) $x_1<x<x_2$, and 2) $x_2\leq x\leq L$.

In case 1, we have
\begin{equation}
\begin{aligned}
&U_n(2,s_{-n})-U_n(1,s_{-n})\\
&=k_p d (p_1-p_2)+k_q(q_1(|A_1|)-q_2(|A_2|))+k_l(2x-x_1-x_2)\\
&> k_p d (p_1-p_2) +k_q(q_1(|A_1|)-q_2(|A_2|))+k_l(x_1-x_2)\\
&=0
\end{aligned}
\end{equation}

In case 2, we have
\begin{equation}
\begin{aligned}
&U_n(2,s_{-n})-U_n(1,s_{-n})\\
&=k_p d (p_1-p_2)+k_q(q_1(|A_1|)-q_2(|A_2|))+k_l(x_2-x_1)\\
&>k_p d (p_1-p_2)+k_q(q_1(|A_1|)-q_2(|A_2|))+k_l(x_1-x_2)\\
&=0
\end{aligned}
\end{equation}

In both cases, selecting station 2 is better than station 1. Therefore, all the PEVs on the right side of station 1 have no incentives to unilaterally change its strategy.

Next, we consider the PEVs on the left side of station 1. When a PEV $m$ at $y$ ($-L\leq y\leq x_1$) adopts the strategy given by (6), we have
\begin{equation}
U_m(1,s_{-m})=-k_pdp_1-k_q q_1(|A_1|)-k_l(x_1-y)
\end{equation}
and
\begin{equation}
U_m(2,s_{-m})=-k_pdp_2-k_q q_2(|A_2|)-k_l(x_2-y)
\end{equation}
Since $\omega_1$ solves \eqref{probability_indifference}, we have $U_m(1,s_{-m})=U_m(2,s_{-m})$.
If PEV $m$ take the strategy given by (6), its payoff is $=\omega_1 U_m(1,s_{-m})+(1-\omega_1)U_m(2,s_{-m})$.

If PEV $m$ takes a different mixed strategy $s^{'}_m=(\omega^{'}_1,1-\omega^{'}_1)$, its payoff is
\begin{equation}
\begin{aligned}
U_m(s^{'}_m,s_{-m})&=\omega^{'}_1 U_m(1,s_{-m})+(1-\omega^{'}_1)U_m(2,s_{-m})\\
&=U_m(1,s_{-m})=U_m(2,s_{-m})\\
&=\omega_1 U_m(1,s_{-m})+(1-\omega_1)U_m(2,s_{-m})
\end{aligned}
\end{equation}
The above equation indicates that PEV $m$ cannot benefit from deviating from the mixed strategy $(\omega_1,1-\omega_1)$ unilaterally.

In all, we can conclude that the strategy profile corresponding to (6) is an NE. Similar to the proof of Theorem 1, we can prove that the NE is unique.

\section{Proof of Theorem 3}
Here, we consider the case of $p_1-p_2 \in [\theta^R_2,p_{\max}-p_{\min}]$.  The proof for $p_1-p_2 \in [p_{\min}-p_{\max},\theta^L_2]$ is similar and hence is omitted.

Consider a PEV $n$ at $x \in [-L,L]$. We have
\begin{equation}
\begin{aligned}
&U_n(2,s_{-n})-U_n(1,s_{-n})=k_p d (p_1-p_2)+k_q(q_1(|A_1|)\\
&-q_2(|A_2|))+k_l(|x-x_1|-|x-x_2|)
\end{aligned}
\end{equation}

In the following, we consider three cases.

In case 1, $x_1<x<x_2$ holds. We have
\begin{equation}
\begin{aligned}
&U_n(2,s_{-n})-U_n(1,s_{-n})=k_p d (p_1-p_2)+k_q(q_1(|A_1|)\\
&-q_2(|A_2|))+k_l(2x-x_1-x_2)\\
&\geq k_p d (p_1-p_2)+k_q(0-q_2(2L))+k_l(2x-x_1-x_2)\\
&\geq k_p d \theta_2^R -k_q q_2(2L)-k_1(x_2-x_1)=0
\end{aligned}
\end{equation}

In case 2, $-L\leq x \leq x_1$ holds. We have
\begin{equation}
\begin{aligned}
&U_n(2,s_{-n})-U_n(1,s_{-n})=k_p d (p_1-p_2)+k_q(q_1(|A_1|)\\
&-q_2(|A_2|))+k_l(x_1-x_2)\\
&\geq k_p d (p_1-p_2)+k_q(0-q_2(2L))+k_l(x_1-x_2)\\
&\geq k_p d \theta_2^R -k_q q_2(2L)-k_1(x_2-x_1)=0
\end{aligned}
\end{equation}

In case 3, $x_2\leq x \leq L$ holds. We have
\begin{equation}
\begin{aligned}
&U_n(2,s_{-n})-U_n(1,s_{-n})=k_p d (p_1-p_2)+k_q(q_1(|A_1|)\\
&-q_2(|A_2|))+k_l(x_2-x_1)\\
&\geq k_p d (p_1-p_2)+k_q(0-q_2(2L))+k_l(x_2-x_1)\\
&\geq k_p d \theta_2^R -k_q q_2(2L)+k_1(x_2-x_1)\\
&>k_p d \theta_2^R -k_q q_2(2L)-k_1(x_2-x_1)=0
\end{aligned}
\end{equation}

It can be seen that selecting station 2 is always the better strategy of all the PEVs. Hence, the NE strategy of each PEV is 2.
In addition, it is clear that there exists no other NE in the game.

\section{Extension for multiple charging stations}
\begin{figure}[h£¡]
\centering
        \scalebox{1}[1]{\includegraphics {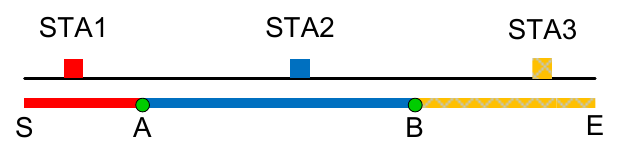}}
 \caption{Two indifference points.}
\end{figure}

\begin{figure}[h£¡]
\centering
        \scalebox{1}[1]{\includegraphics {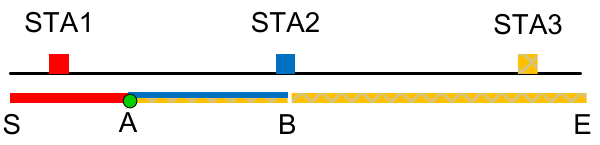}}
 \caption{One unique indifference point.}
\end{figure}
In Section IV, we characterize the conditions for the existence of an indifference point between two charging stations. Now we consider a one-dimensional system with \emph{three} charging stations. A similar analysis can be generalized to the case with an \emph{arbitrary} number of stations in a one-dimensional system.

As shown in Fig. 17, it is possible to have two indifference points in this case, denoted by A and B. At point A, a PEV is indifferent from selecting between stations 1 and 2, but is unwilling to select station 3. At point B, a PEV is indifferent from selecting between stations 2 and 3, but is unwilling to select station 1. Under an NE characterized by these indifference points, the PEVs in the line segment S-A select station 1, those in the line segment A-B select station 2, and those in the line segment B-E select station 3.

Another interesting case is shown in Fig. 18, where only one indifference point A exists. At point A, a PEV is indifferent from selecting among all three stations. Accordingly, the PEVs in the line segment S-A prefer station 1 under the corresponding NE. On the other hand, the PEVs in the line segment A-B take a mixed-strategy, which selects station 2 with a probability $\omega$ and station 3 with a probability $1-\omega$. The PEVs in the line segment B-E prefer station 3 at the NE.

\section{Extension for a two-dimensional system}
\begin{figure}[h!]
\centering
        \scalebox{1}[1]{\includegraphics {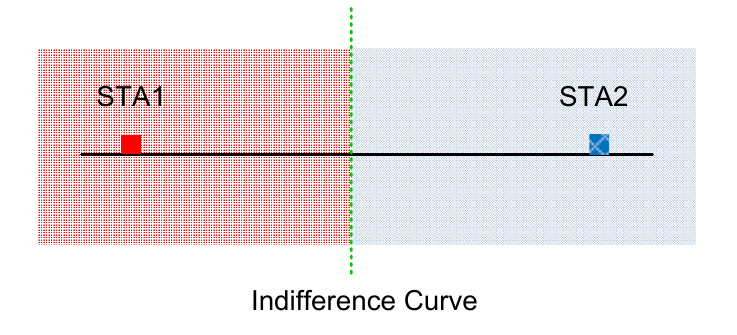}}
 \caption{A indifference line.}
\end{figure}

\begin{figure}[h!]
\centering
        \scalebox{0.5}[0.5]{\includegraphics {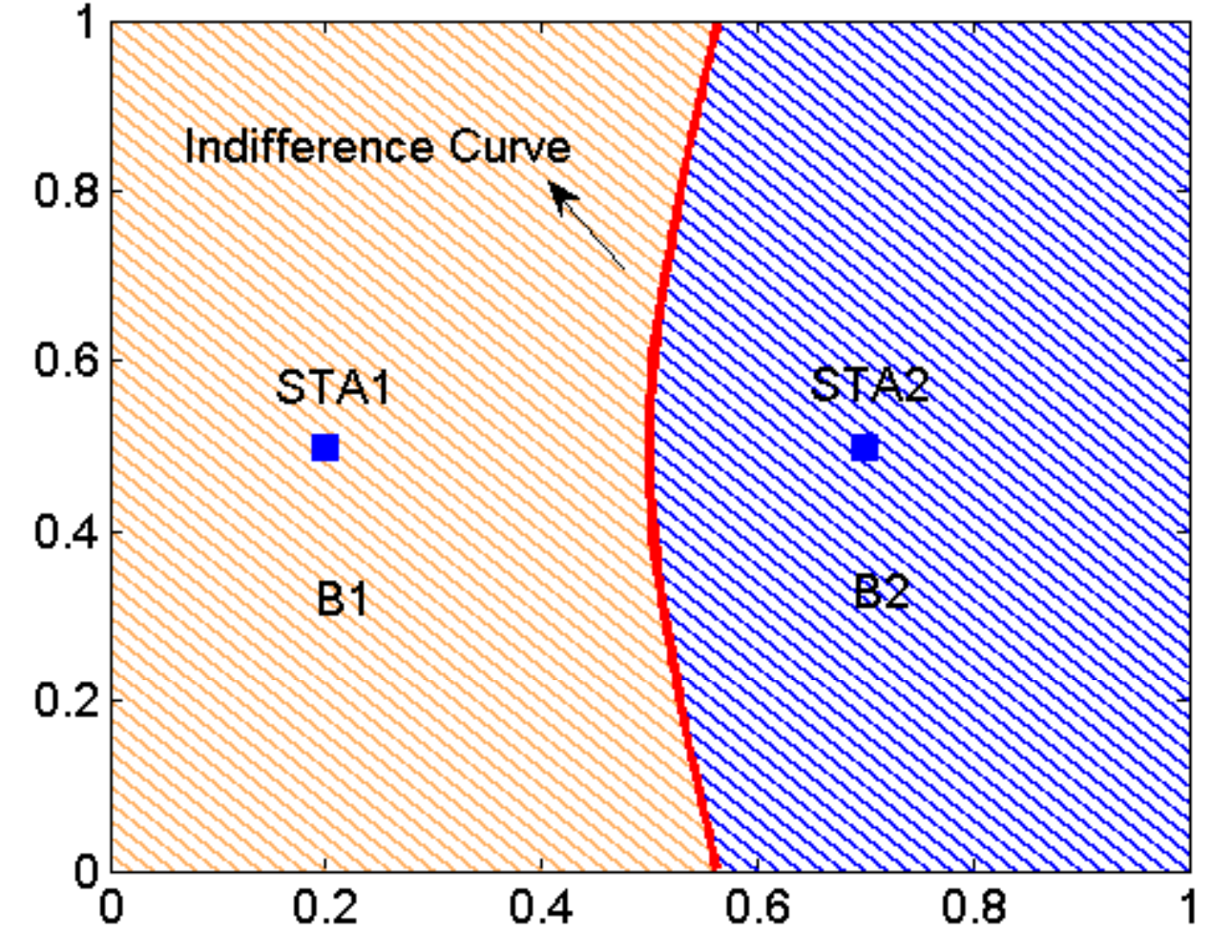}}
 \caption{A indifference curve.}
\end{figure}

Now we extend our model to a \emph{two-dimensional} system. For simplicity, we first consider two \emph{homogeneous} stations (i.e., $\mu_1=\mu_2$, $\sigma_1=\sigma_2$, $k_1=k_2$ and $p_1=p_2$) symmetrically located in a two-dimensional system. In this case, there exists one \emph{indifference line} (instead of indifference point) which goes through the midpoint between two stations, as show in Fig. 19. At the corresponding NE, all the PEVs on the left side of the indifference line will select station 1, and the others select station 2.

When the stations are asymmetrical, we will end up with a nonlinear \emph{indifference curve}. We denote by $(x_1,y_1)$ and $(x_2,y_2)$ the positions of charging stations $1$ and $2$, respectively. Let $(x_n,y_n)$ be the position of a PEV on the indifference curve, which satisfies the following condition:
\begin{equation}
k_l l_{n,1} + k_q q_1 + k_p d p_1 = k_l l_{n,2} + k_q q_2 + k_p d p_2.
\end{equation}
Here, $l_{n,i}=\sqrt{(x_n-x_1)^2+(y_n-y_1)^2}$ denotes the distance between the PEV and station $i$, and $q_i$ represents the queueing delay suffered at station $i\in \{1,2\}$. According to the M/G/k queueing theory, we have
\begin{equation}\label{waiting time}
q_i(|B_i|) \approx\frac{|B_{i}|\lambda (\sigma^2_{i}+\frac{1}{\mu_{i}^2})\rho_{i}^{k_{i}-1}}{2(k_{i}-1)!(k_{i}-\rho_{i})^2\large[\sum_{m=0}^{k_{i}-1}\frac{\rho_{i}^m}{m!} +\frac{\rho_{i}^{k_{i}}}{(k_{i}-1)!(k_{i}-\rho_{i})}\large]},
\end{equation}
where $\rho_i=\frac{|B_i|\lambda}{\mu_i}$ and $|B_i|$ represents the \emph{area} of the region served by station $i$. For illustration, we consider a two-dimensional system shown in Fig. 20. The coverage area of this system is $1.0\times 1.0$, and the positions of the two stations are $(0.2, 0.5)$ and $(0.7, 0.5)$, respectively. Suppose that the price of station 1 is lower than that of station 2. The red curve between two stations is the indifference curve. The indifference curve splits the system into two regions (i.e., $B_1$ and $B_2$), and $|B_1|$ (or $|B_2|$) is the area of the region containing $STA_1$ (or $STA_2$).

\section{Proof of Theorem 6}

Consider two prices $p^1_i$ and $p^2_i$ ($p^1_i<p^2_i$). Since $\mathcal{B}_j(\cdot)$ is increasing, we have $\mathcal{B}_j(p^1_i)\leq \mathcal{B}_j(p^2_i)$ and $\mathcal{B}_i(\mathcal{B}_j(p^1_i))\leq \mathcal{B}_i(\mathcal{B}_j(p^2_i))$. For convenience, we use $\mathcal{F}(\cdot)$ to refer to $\mathcal{B}_i(\mathcal{B}_j(\cdot)$. Clearly, $\mathcal{F}(\cdot)$ is non-decreasing. Due to $\mathcal{F}(a)\geq a$ and $\mathcal{F}(b)\leq b$, we know that $a\leq \mathcal{F}(x)\leq b$ if $a<x<b$. According to Theorem 12.5 of \cite{game book}, there exists one fixed point satisfying $\mathcal{B}_j(\mathcal{B}_i(p_j))=p_j$, which indicates the existence of pure-strategy NE.

Now we prove the uniqueness of pure-strategy NE by contradiction. Assume that there exist two NEs, \emph{i.e.}, $(p^a_1,\mathcal{B}_2(p^a_1))$ and $(p^b_1,\mathcal{B}_2(p^b_1))$. Without loss of generality, we suppose $p^a_1>p^b_1$. According to condition 3), we have $\mathcal{B}_2(p^a_1)-p^a_1< \mathcal{B}_2(p^b_1)-p^b_1$. Furthermore, we know $\mathcal{B}_2(p^a_1)\geq \mathcal{B}_2(p^b_1)$ since $p^a_1>p^b_1$. Therefore, we have $\mathcal{B}_1(\mathcal{B}_2(p^a_1))-\mathcal{B}_2(p^a_1)\leq \mathcal{B}_1(\mathcal{B}_2(p^b_1))-\mathcal{B}_1(p^b_1)$. Since $(p^a_1,\mathcal{B}_2(p^a_1))$ is an NE, $\mathcal{B}_1(\mathcal{B}_2(p^a_1))-p^a_1=0$ holds. Then we have
\begin{equation}
\begin{aligned}
\mathcal{B}_1(\mathcal{B}_2(p^b_1))-p^b_1&=\mathcal{B}_1(\mathcal{B}_2(p^b_1))-\mathcal{B}_2(p^b_1)+\mathcal{B}_2(p^b_1)-p^b_1\\
&> \mathcal{B}_1(\mathcal{B}_2(p^a_1))-\mathcal{B}_2(p^a_1)+\mathcal{B}_2(p^a_1)-p^a_1=0,
\end{aligned}
\end{equation}
which contradicts the assumption that $(p^b_1,p^b_2)$ is an NE.

\section{Proof of Proposition 1}
For simplicity, we only show the proof for $p_i>p_i^*$. The proof for $p_i<p_i^*$ is similar and omitted here.
Since $\mathcal{B}_i(p_j)$ ($i=1,2$) is increasing and $B_i(p_j)-p_j$ ($\forall i\neq j$) is strictly decreasing, we have
\begin{equation}
\begin{aligned}
\Theta_i(p_i)&=\mathcal{B}_i(\mathcal{B}_j(p_i))-p_i\\
&=\mathcal{B}_i(\mathcal{B}_j(p_i))-\mathcal{B}_j(p_i)+\mathcal{B}_j(p_i)-p_i\\
&< \mathcal{B}_i(\mathcal{B}_j(p_i))-\mathcal{B}_j(p_i)+\mathcal{B}_j(p_i^*)-p_i^*\\
&\leq \mathcal{B}_i(\mathcal{B}_j(p_i^*))-\mathcal{B}_j(p_i^*)+\mathcal{B}_j(p_i^*)-p_i^*\\
&=0.
\end{aligned}
\end{equation}
It completes the proof.

\section{Proof of Theorem 7}
\begin{figure}[h]\label{convergence_proof}
\centering
        \scalebox{0.8}[0.8]{\includegraphics {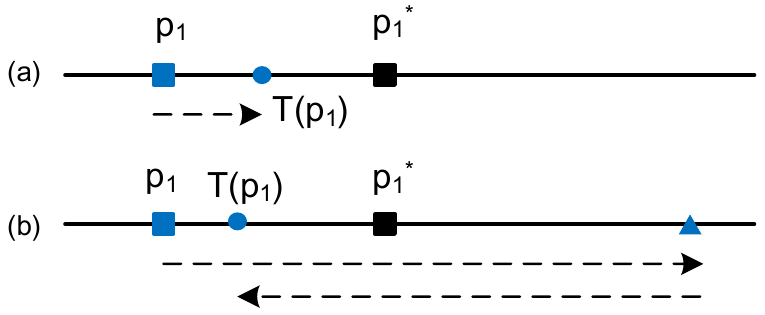}}
        \caption{Mapping of DSSA.}
\end{figure}
For simplicity, we only consider the case where $i=1$ and $\alpha$ is sufficient large~\footnote{Our proof can be easily extended to the case where $\alpha$ is small. In this case, we only need to consider a different starting point which is enough close to the equilibrium point.}. For any non-equilibrium point $p_1(t)$, there exist two cases: 1) $p_1(t)<p^*_1$, and 2) $p_1(t)>p^*_1$. In the following, we only consider case 1, and case 2 can be analyzed in a similar fashion. Let $\mathcal{P}_1(\tau)=p_1(t)$. As shown in Fig. 21, we can define a mapping as
\begin{equation}
\mathcal{T}(\mathcal{P}_1(\tau)=
   \begin{cases}
   p_1(t+1), & \mbox{if $\Theta_1(p_1(\tau+1)) > 0$},\\
   p_1(t+2), & \mbox{if $\Theta_1(p_1(\tau+1)) < 0$}, \\
   \end{cases}
\end{equation}
Clearly, our iterative algorithm DSSA can be characterized by $\mathcal{P}_1(\tau+1)=\mathcal{T}(\mathcal{P}_1(\tau))$. Now we consider two cases.

(1) If $\mathcal{T}(\mathcal{P}_1(\tau)=p_1(t+1)$, we have
\begin{equation}
|\mathcal{T}(\mathcal{P}_1(\tau))-p^*_1|=p_1^*-p_1(t)-\delta(t)< |p_1^*-p_1(t)|=|p_1^*-\mathcal{P}_1(\tau)|
\end{equation}

(2) If $\mathcal{T}(\mathcal{P}_1(\tau)=p_1(t+2)$, we have
\begin{equation}
\begin{aligned}
|\mathcal{T}(\mathcal{P}_1(\tau))-p^*_1|&=p^*_1-(p_1+\delta(t)-\alpha \delta(t))\\<|p_1^*-p_1(t)|=|p_1^*-\mathcal{P}_1(\tau)|
\end{aligned}
\end{equation}

In both cases, the distance between the candidate solution and the equilibrium is shortened by the mapping. There exist a positive constant $\gamma$ ($0\leq \gamma<1$) satisfying $|\mathcal{T}(\mathcal{P}_1(\tau))-p^*_1|\leq \gamma|p_1^*-\mathcal{P}_1(\tau)|$ ($\forall \tau>0$). It indicates $\mathcal{T}$ is a \emph{pseudocontraction} \emph{mapping}~\cite{Parallel}. According to Proposition 1.2 of \cite{Parallel}, DSSA converges to $p^*_1$. Further, we know the convergence is Q-linear \cite{numerical optimization}.

\end{appendices}

\begin{IEEEbiography}[{\includegraphics[width=1in,height=1.25in,clip,keepaspectratio]{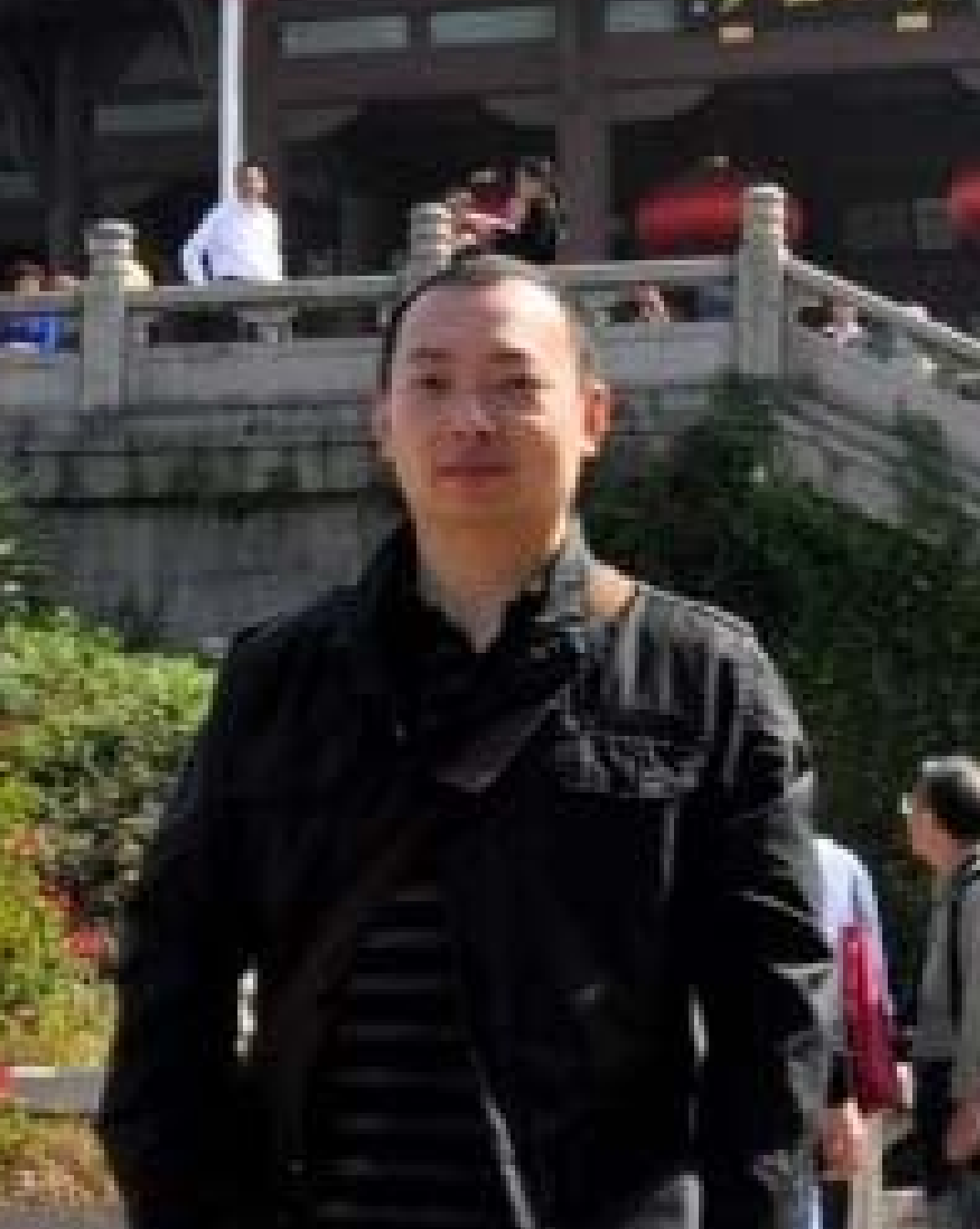}}]{Wei Yuan}
received the B.S. degree in electronic engineering from Wuhan University, China, in 1999, and the Ph.D. degree in electronic engineering from the University of Science and
Technology of China, Hefei, in 2006. He is currently an associate professor in the School of Electronic Information and Communications at Huazhong University of Science and
Technology, China. His current research interests include wireless networks, EV charging networks, and the applications of optimization, game theory and machine learning.
\end{IEEEbiography}

\begin{IEEEbiography}[{\includegraphics[width=1in,height=1.25in,clip,keepaspectratio]{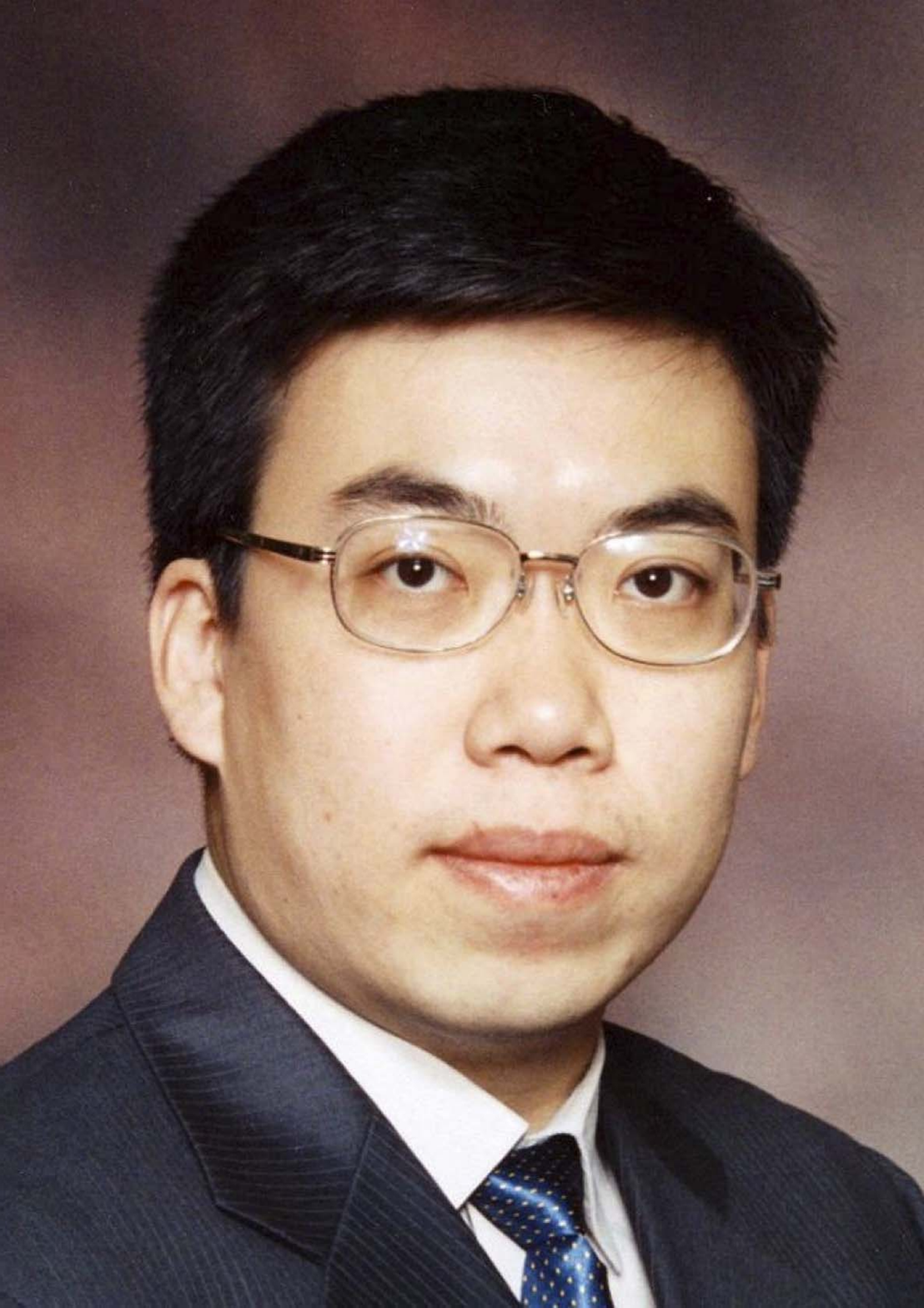}}]{Jianwei Huang}
Jianwei Huang (S'01-M'06-SM'11-F'16) is an Associate Professor and Director of the Network Communications and Economics Lab (ncel.ie.cuhk.edu.hk), in the Department of Information Engineering at the Chinese University of Hong Kong. He received the Ph.D. degree from Northwestern University in 2005, and worked as a Postdoc Research Associate in Princeton during 2005-2007. He is the co-recipient of 8 international Best Paper Awards, including IEEE Marconi Prize Paper Award in Wireless Communications in 2011. He has co-authored four books: ``Wireless Network Pricing," ``Monotonic Optimization in Communication and Networking Systems,"  ``Cognitive Mobile Virtual Network Operator Games," and ``Social Cognitive Radio Networks". He has served as an Editor of several top IEEE Communications journals, including JSAC, TWC, and TCCN, and a TPC Chair of many international conferences. He is the Vice Chair of IEEE ComSoc Cognitive Network Technical Committee and the Past Chair of IEEE ComSoc Multimedia Communications Technical Committee. He is a Fellow of IEEE (Class of 2016) and a Distinguished Lecturer of IEEE Communications Society.
\end{IEEEbiography}

\begin{IEEEbiography}[{\includegraphics[width=1in,height=1.25in,clip,keepaspectratio]{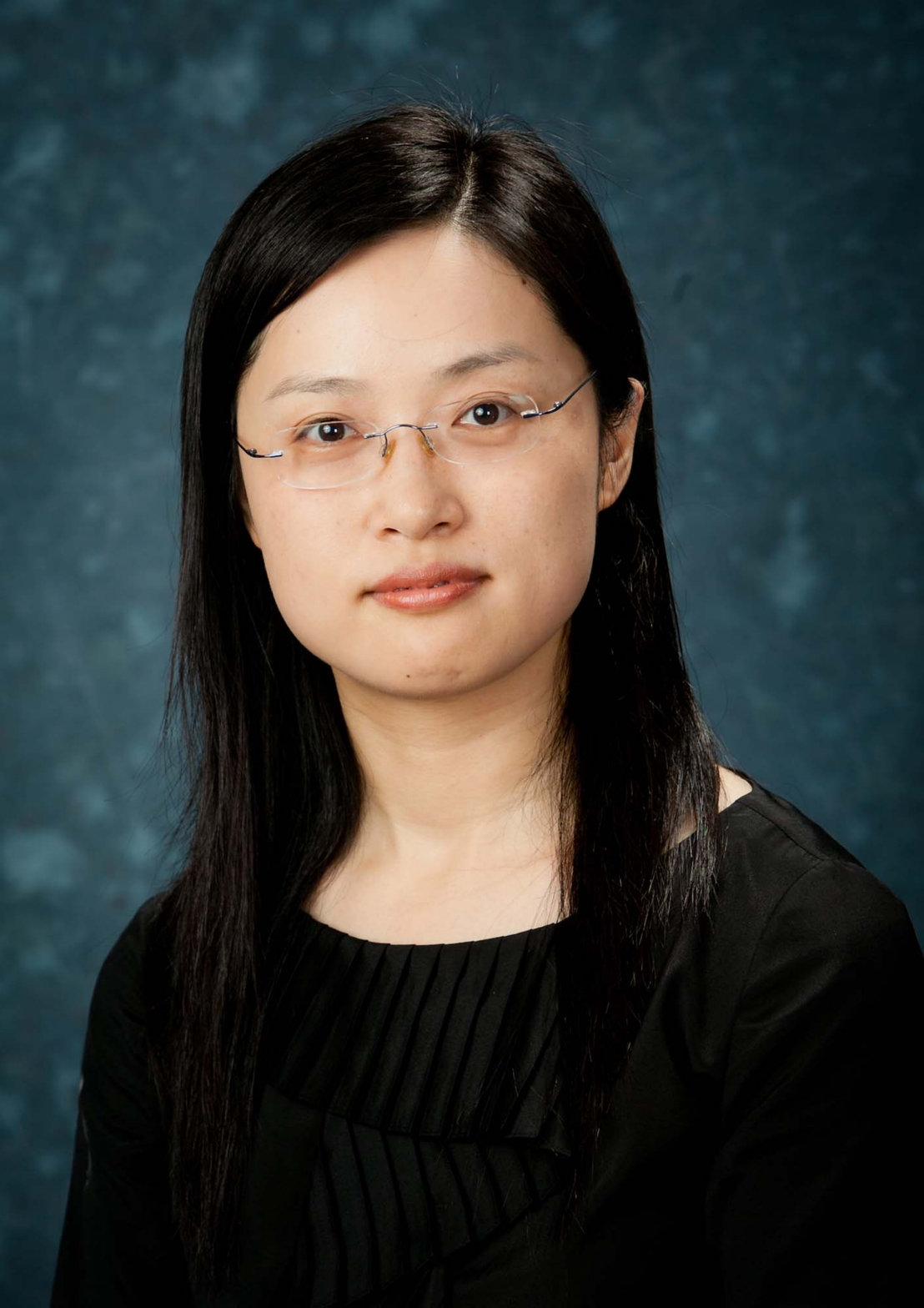}}]{Ying Jun (Angela) Zhang}
Ying Jun (Angela) Zhang (S'00-M'05-SM'11) received the B.Eng. degree in electronic engineering from Fudan University, Shanghai, China, in 2000, and the Ph.D. degree in electrical and electronic engineering from the Hong Kong University of Science and Technology, Hong Kong, in 2004.

Since 2005, she has been with the Department of Information Engineering, The Chinese University of Hong Kong, where she is currently an Associate Professor. During the summers of 2007 and 2009, she was with Wireless Communications and Network Science Laboratory at Massachusetts Institute of Technology (MIT). Her current research interests are mainly focused on wireless communications systems and smart power systems, in particular optimization techniques for such systems.

She is an Executive Editor of IEEE TRANSACTIONS ON WIRELESS COMMUNICATIONS. She is also an Associate Editor of IEEE TRANSACTIONS ON COMMUNICATIONS. Previously, she served many years as an Associate Editor of IEEE TRANSACTIONS ON WIRELESS COMMUNICATIONS and Security and Communications Networks (Wiley), and a Guest Editor of a Feature Topic in IEEE COMMUNICATIONS MAGAZINE. She has served as a Workshop Chair of IEEE ICCC 2014 and 2013, TPC Vice Chair of Wireless Networks and Security Track of IEEE VTC 2014, TPC Vice-Chair of Wireless Communications Track of IEEE CCNC 2013, TPC Co-Chair of Wireless Communications Symposium of IEEE GLOBECOM 2012, Publication Chair of IEEE TTM 2011, TPC Co-Chair of Communication Theory Symposium of IEEE ICC 2009, Track Chair of ICCCN 2007, and Publicity Chair of IEEE MASS 2007. She was a Co-Chair of IEEE ComSoc Multimedia Communications Technical Committee and the IEEE Communication Society GOLD Coordinator.

Dr. Zhang is a co-recipient of 2014 IEEE ComSoc APB Outstanding Paper Award, 2013 IEEE SmartgridComm Best Paper Award, and 2011 IEEE Mar- coni Prize Paper Award on Wireless Communications. She received the Young Researcher Award from The Chinese University of Hong Kong in 2011. As the only winner from Engineering Science, she has won the Hong Kong Young Scientist Award 2006, conferred by the Hong Kong Institution of Science.
\end{IEEEbiography}

\end{document}